\documentclass[a4paper,fleqn,usenatbib,useAMS]{mnras}
\usepackage{graphicx}
\usepackage{amsmath}
\usepackage{amssymb}
\usepackage{pdflscape}
\usepackage{multicol}
\usepackage[T1]{fontenc}
\usepackage{ae,aecompl}
\usepackage{txfonts}

\title[Near-Infrared Photometry of Globular Clusters]{Near-Infrared Photometry
  of Globular Clusters Towards the Galactic Bulge: Observations and Photometric Metallicity Indicators}
\author[R. E. Cohen et al.]{
Roger E. Cohen$^{1}$\thanks{E-mail: rcohen@astro-udec.cl},
Christian Moni Bidin$^{2}$,
Francesco Mauro$^{3,1}$,
Charles Bonatto$^{4}$\\
\newauthor{and Douglas Geisler$^{1}$}
\\
$^{1}$Departamento de Astronom\'{i}a, Universidad de Concepci\'{o}n, Casilla 160-C, Concepci\'{o}n, Chile\\
$^{2}$Instituto de Astronom\'ia, Universidad Cat\'olica del Norte, Av. Angamos 0610, Casilla 1280, Antofagasta, Chile\\
$^{3}$Instituto Milenio de Astrof\'isica, Santiago, Chile\\
$^{4}$Departamento de Astronomia, Universidade Federal do Rio Grande do Sul, Av. Bento Gon\c{c}alves 9500 Porto Alegre 91501-970, RS, Brazil} 

\date{Accepted XXX. Received YYY; in original form ZZZ}

\pubyear{2016}

\begin{document}
\label{firstpage}
\pagerange{\pageref{firstpage}--\pageref{lastpage}}
\maketitle

\begin{abstract}
We present wide field $JHK_{S}$ photometry of 16 Galactic
globular clusters located towards the Galactic bulge, calibrated on the
2MASS photometric system.  Differential reddening 
corrections and statistical field star decontamination 
are employed for all of these clusters before fitting fiducial
sequences to the cluster red giant branches (RGBs).
Observed values and uncertainties are reported
for several photometric features, 
including the magnitude of the RGB bump, tip, the horizontal branch (HB)
and the slope of the upper RGB.
The latest spectroscopically determined chemical abundances are used
to build distance- and reddening-independent relations between 
observed photometric features and cluster
metallicity, optimizing the sample size and metallicity baseline of these 
relations by supplementing our sample with
results from the literature.  
We find that the magnitude difference between the HB and the RGB bump can be
used to predict metallicities, in terms of both iron abundance $\mathit{[Fe/H]}$ and 
global metallicity $\mathit{[M/H]}$, with a precision of better than 0.1 dex in all 
three near-IR bandpasses for relatively metal-rich ($[M/H]$$\gtrsim$-1) clusters.  
Meanwhile, both the slope of the upper RGB and the magnitude
difference between the RGB tip and bump are useful metallicity indicators over
the entire sampled metallicity range (-2$\lesssim$$[M/H]$$\lesssim$0) with a 
precision of 0.2 dex or better, despite model predictions that the RGB slope may become unreliable at high (near-solar) metallicities.    
Our results agree with previous calibrations in light of the relevant 
uncertainties, and we discuss implications for clusters 
with controversial metallicities
as well as directions for further investigation.  

\end{abstract}


\section{Introduction}

Galactic globular clusters (GGCs) play a crucial role in constraining stellar
evolutionary models as well as Galactic chemical evolution.  Recently, many of
these clusters have been the subject of large-scale photometric surveys
using deep, high resolution multi-colour space-based observations 
\citep{piotto,sarajedini,piotto14}.
However, GGCs located towards the Galactic bulge, despite their importance as
the most metal-rich (and in some cases, massive) members of the GGC system, 
have been generally
excluded from these surveys due to severe total and differential
extinction at optical wavelengths.
For
this reason, infrared wavelengths, where the effects of extinction are greatly
reduced ($A_{K}$$\sim$$0.12A_{V}$; \citealt{casagrande}), are ideal for 
photometric investigations of such clusters.  

The \textit{Vista Variables in the Via Lactea} (VVV), an ESO public survey,
has observed a 562 sq.~degree field including the Galactic bulge
and a portion of the disk in $YZJHK_{S}$ filters down to $K_{S}$$\sim$20, and
thus presents an ideal opportunity to study the 
GGCs located in the survey area.  Since the advent of near-IR arrays, a wealth of effort has been 
devoted to studying GGCs in the near-infrared
largely by Valenti, Ferraro and collaborators
(e.g.~\citealt{ferraro00,v04obs,v04abs}, Valenti et al.~2010, 
hereafter \citealt{v10}; 
also see \citealt{chun} and references therein), in addition to the earlier
studies of \citet{chobump} and \citet{jura2mass} which employed photometry 
from the Two Micron All Sky Survey (2MASS; \citealt{skrutskie}).  
An important goal of these investigations was the construction of relations 
between observable features in
cluster near-IR colour-magnitude diagrams (CMDs) and their chemical abundances,
as these relations can then be applied to obtain photometric metallicity
estimates.    
With an eye towards future application for distant and/or heavily
extincted stellar systems, we revisit these calibrations.  This is 
advantageous
in light of not only the quality of the VVV
photometry, but more importantly its wide-field nature, facilitating
a statistical assessment of contamination by field stars (see Sect.~\ref{decontbigsect}), leveraged together
with improved spectroscopic abundances (see Sect.~\ref{badfehsect}) 
and reddening maps (e.g.~\citealt{javier12,gonzalez,cohen6544}; see Sect.~\ref{diffredsect}).  
Here we analyse an
initial subset of 
GGCs within the VVV survey area which have 
spectroscopically measured $\mathit{[Fe/H]}$ values, with the goal of 
constructing updated distance- and reddening-independent relations
between photometric features observable on the cluster giant and horizontal
branches and their metallicities.  The resulting relations between distance-
and reddening-independent photometric features measured from near-IR cluster
CMDs versus cluster metallicities are further optimized by concatenating the
results presented here with those available in the literature.  

In the next section, we present the details of our observations and data
processing, including corrections for differential reddening and field star
contamination, and the resulting cluster CMDs.  In Sect.~3, 
we describe our methodology for measuring 
cluster photometric features as well as their uncertainties, and in Sect.~4
we use these measurements, along with literature values, 
to construct relations which can be used to estimate
metallicities of old stellar populations photometrically.
In the final section we summarize our results, discussing implications
for clusters with controversial metallicity values.  

\section{Data Processing}

\subsection{Target Cluster Selection }
\label{metsect}

There are 36 Galactic globular clusters (GGCs) presently known in the area
covered by the VVV survey according to the catalog of Harris (1996, 2010 revision, hereafter \citealt{h96}), plus one candidate 
discovered as a result of this survey
\citep[VVV CL001,][]{minniti}.
We aim to derive
relations between observed photometric
parameters on the cluster red giant branches (RGBs), where the most
IR-bright cluster members lie, and cluster metallicities 
(in terms of both $\mathit{[Fe/H]}$ and $\mathit{[M/H]}$), so  
we have selected a subset of the GGCs in the VVV survey area 
which all have spectroscopically
measured $\mathit{[Fe/H]}$ values.  
To restrict our sample to only those clusters with high quality $\mathit{[Fe/H]}$
measurements, we consider only clusters with a value of "1" in the last column 
of Table A.1 in Carretta et al.~(2009, hereafter \citealt{c09}), and add two
clusters (NGC 6380 and M 28=NGC 6626) with recent spectroscopic $\mathit{[Fe/H]}$ values based on
CaII triplet equivalent widths (\citealt{saviane}; Mauro et
al.~2014, hereafter \citealt{mauro14}), comprising a sample of 17 GGCs from
VVV including photometry of NGC 6544 described in \citet{cohen6544}.
We return to the issue of various spectroscopic
metallicities for the target clusters in Sect.~\ref{badfehsect}, and 
the use of literature measurements for additional clusters is discussed in Sect.~\ref{baselinesect}.

\subsection{Photometry}
\begin{figure*}
\includegraphics[width=0.95\textwidth]{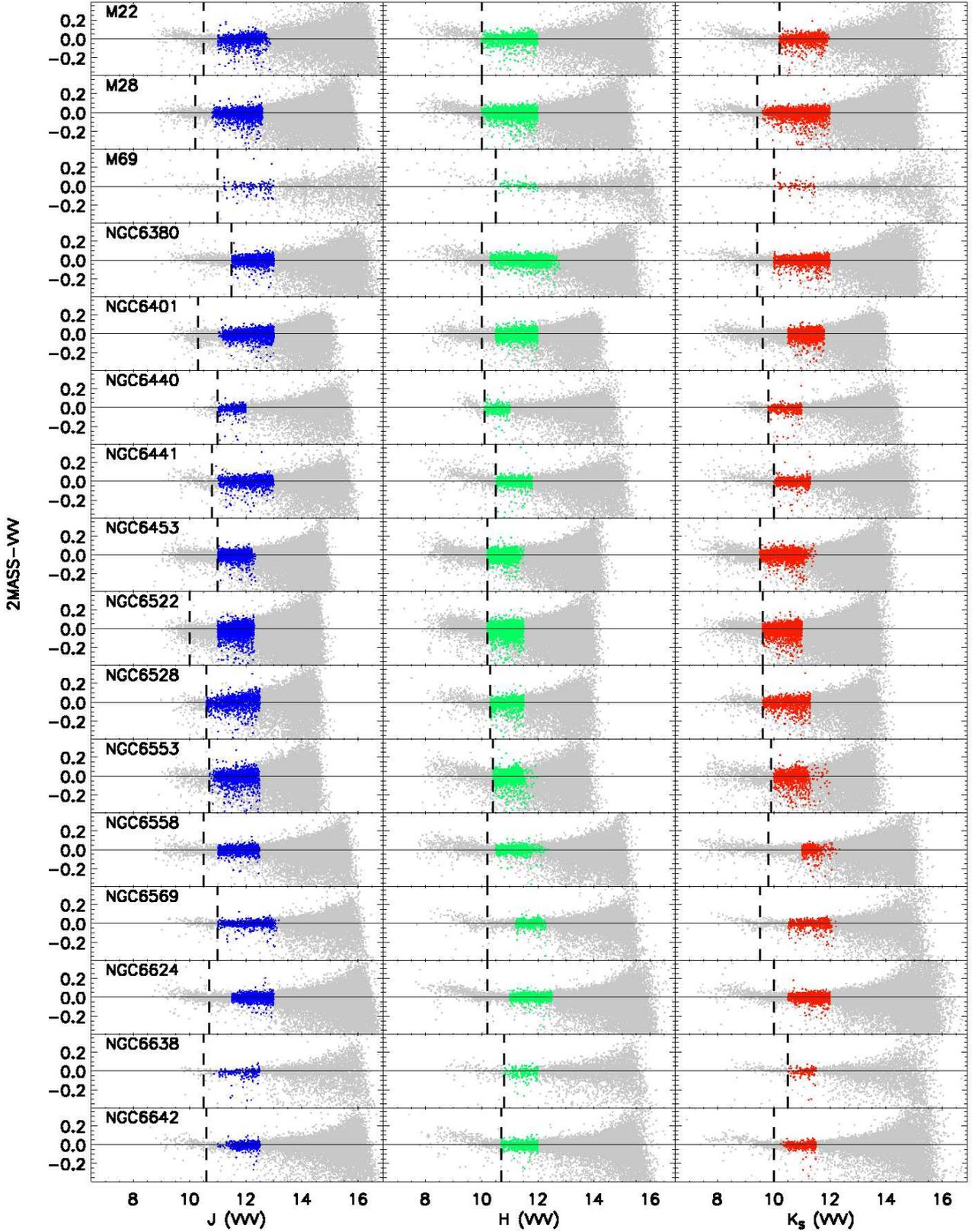}
\caption{A comparison between our calibrated magnitudes and those from 2MASS
for all of our target clusters.  Each cluster is shown as a row of
three plots, illustrating the difference between VVV and 2MASS as a function
of (left to right) VVV $J$, $H$ and $K_{S}$ magnitude.  In each plot, the grey
points represent all stars matched between VVV and 2MASS, while the coloured
points represent the stars used for calibration.  The solid horizontal
line represents equality, while the dashed vertical line indicates the VVV
saturation limit above which photometry from 2MASS was employed.}
\label{comp2mass}
\end{figure*}

The images which we employ were obtained as part of the VVV survey
using the 4.1m Visible and Infrared Survey Telescope for Astronomy (VISTA),
equipped with the VIRCAM (VISTA InfraRed Camera) instrument \citep{emerson}.  
The VIRCAM detector consists
of a 4$\times$4 array of chips, each with 2048x2048 pixels and a pixel
scale of 0.339$\arcsec$ per pixel.  A description of the survey can be found 
in \citet{minniti}, with further details regarding the survey strategy and
data products in \citet{saito10}.  Information regarding the first
data release, including products which we employ here, 
is given in \citet{saito}.  
Point-spread-function fitting (PSF) photometry is performed on VVV images
obtained from the Cambridge Astronomical Survey Unit (CASU)\footnote{Images and aperture photometry catalogs from VVV data releases are publicly
  available through the ESO archive, and CASU is located at
  http://casu.ast.cam.ac.ck} via the iterative
usage of the DAOPHOT/ALLFRAME suite \citep{stet87,stet94} identically
to previous studies
\citep[e.g.][]{maurohb,cohen6544}.  This PSF photometry pipeline has been
customized to operate on preprocessed, stacked VVV images produced by CASU, 
and the reader is referred
to \citet{mauropipe} for a detailed description of the PSF
photometry pipeline and comparisons with other data reduction
techniques and products.  
We have chosen to perform photometric and astrometric
calibration of the resulting catalogs 
to 2MASS 
for two reasons.
First, because our photometry becomes saturated
below the tip of the RGBs of all of our target clusters, merging our photometric
catalogs with 2MASS is necessary in order to construct fiducial sequences and 
luminosity functions (LFs)  
over the entire luminosity range of the cluster RGBs and 
measure photometric features 
(described in Sect.~\ref{anasect}).
Second, by performing our analysis in the 2MASS photometric system, our results
may be directly compared and/or combined with 
previous near-IR studies, the majority of which have been calibrated
to 2MASS as well 
\citep{v04obs,v10,chun,cohen6544,cohenispi}.
To calibrate our photometry and astrometry to the 2MASS $JHK_{S}$ system, 
a magnitude range is
selected among the stars matched between VVV and the 2MASS point source
catalog in which both datasets show good agreement with minimal scatter, 
avoiding stars
which are sufficiently faint so as to be unduly affected by crowding
and/or large photometric errors in 2MASS.  Additionally, stars with neighbors
detected within 2.2$\arcsec$ contributing a contaminating flux 
of $\geq$0.03 mag are
rejected from use as local standards \citep[e.g.][]{mauropipe}.
Instrumental magnitudes resulting from PSF photometry are calibrated to the 
2MASS $JHK_{S}$ system (rather than the native VISTA filter system) 
using the classical transformation equations of the form 
$m_{2MASS}-m_{inst} = a + b(J-K_{S})_{2MASS}$,
where $a$ is a photometric zeropoint offset and $b$ is a linear colour term. 
The coefficients $a$ and $b$ are obtained independently for each VIRCAM chip
per image per filter using least squares fitting, 
but using a weighting scheme to downweight discrepant data points\footnote{The algorithm is based
on a series of lectures presented at "V Escola Avancada de Astrofisica" by P.~B.~Stetson,
see \url{http://ned.ipac.caltech.edu/level5/Stetson/Stetson\_contents.html} and 
\url{http://www.cadc.hia.nrc.gc.ca/community/STETSON/homogenous/}} 
rather than a sigma clipping or rejection procedure.  
For the coefficients $a$ and $b$, the values measured in each of the 
three ($J,H,K_{S}$) filters are $a$=(0.62,0.26,-0.52)$\pm$(0.04,0.03,0.06) 
and $b$=(0.03,-0.02,-0.02)$\pm$(0.02,0.02,0.02), compared to median fitting
uncertainties $\leq$0.02 for the offset $a$ and $\leq$0.01 for the colour
term $b$ in all three bandpasses.  
Thus, the resulting photometric 
calibrations have 1$\sigma$ zeropoint uncertainties of $\lesssim$0.02 mag for
all target clusters, and
a star by star comparison between our calibrated photometry and 2MASS in all 
three $JHK_{S}$ filters is shown in Fig. \ref{comp2mass}.  
All stars matched between VVV and 2MASS are shown in grey in each panel of 
Fig.~\ref{comp2mass}, and the subset of these stars used for calibration 
is overplotted.  The vertical dashed line in each
panel of Fig.~\ref{comp2mass} indicates the magnitude at which the VVV 
photometry is unusable due to saturation,
which varies somewhat from cluster to cluster due to
differences in stellar crowding as well as observing conditions.  
For stars which are brighter than this limit in any of the three
$JHK_{S}$ filters, we 
supplement our VVV catalogs with photometry from the 2MASS point source
catalog (PSC).  All colours and magnitudes which we report in this 
study are in the 2MASS photometric system (rather than the native VISTA
system), 
and additional discussions
regarding the calibration of VVV photometry to the 2MASS system can be found in 
\citet{monibidin} and \citet{chene2mass}.
  
Astrometric calibration is performed
to the coordinates given in the 2MASS PSC,
using the world coordinate system information placed in the headers of the 
stacked
VVV images by CASU as an initial guess in order to correct for effects of
geometric distortion.  The resulting astrometry has a
root mean square (rms) precision of $\sim$0.2$\arcsec$ for all target
clusters, in accord with the  
astrometric precision of 2MASS.

\subsection{Comparison With Previous Photometry}

\begin{figure*}
\includegraphics[width=0.95\textwidth]{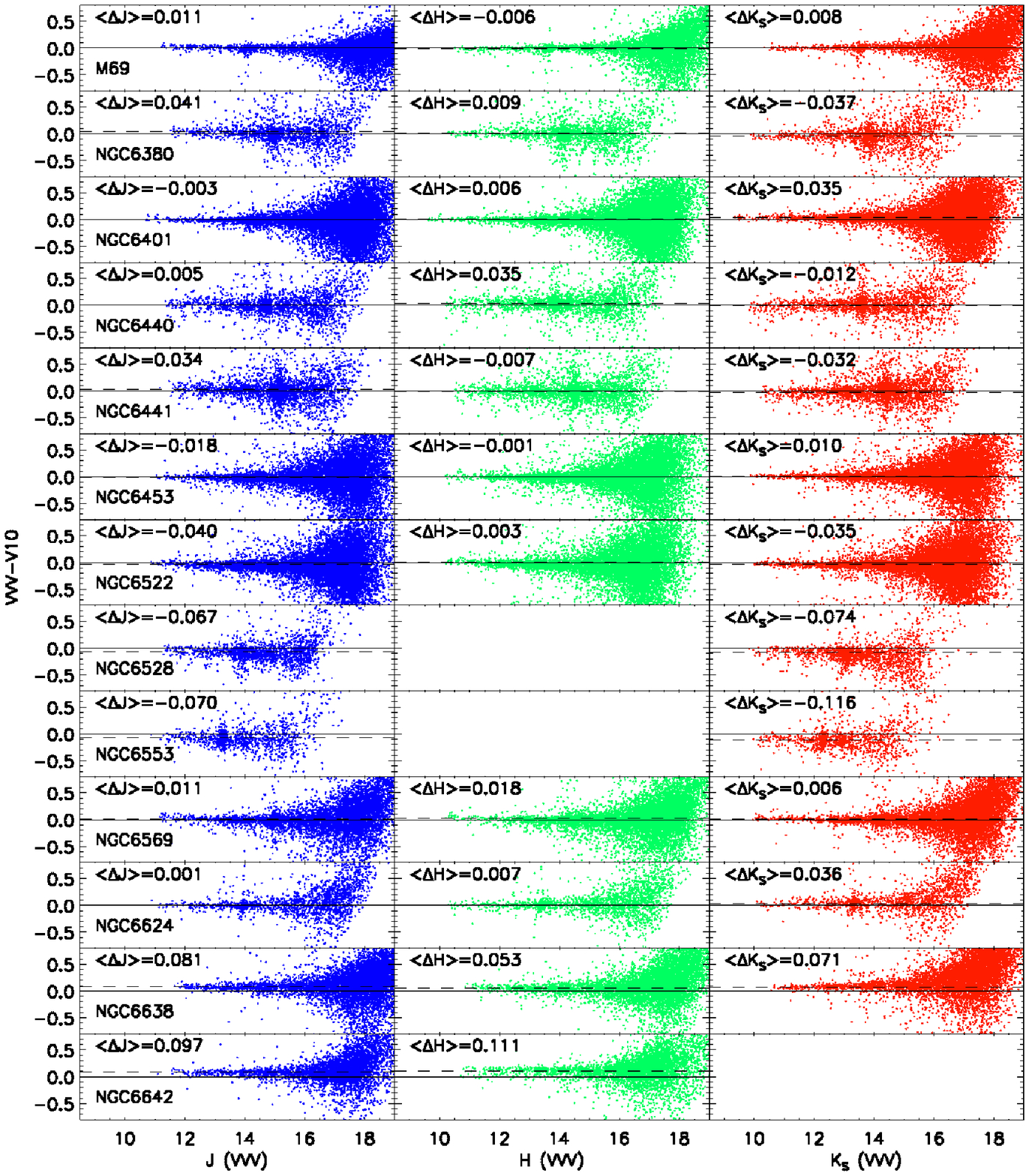}
\caption{A comparison between our photometry and that of \citet{v10}.  Symbols
are as in Fig. \ref{comp2mass} except that the mean magnitude offset is given 
in each plot and shown as a horizontal dashed line.}
\label{compV04}
\end{figure*}

Of our target clusters, 13 of 16 are also included in the compilation of
\citet{v10}\footnote{See the Bulge Globular Cluster Archive
  at http://www.bo.astro.it/$\tilde{ }$GC/ir\_archive}.  We calculate the mean 
magnitude differences in each filter between our photometry and theirs 
using a weighted 2.5$\sigma$ clip in magnitude bins, employing only
unsaturated stars brightward of the observed LF peak.  The
resulting comparisons of magnitude difference as a function of magnitude 
are shown for each cluster in Fig. \ref{compV04}.  Given our
photometric zeropoint uncertainty of $<$0.02 mag and the zeropoint 
uncertainty of 0.05 mag estimated by \citet{v10}, the two studies, 
having both been calibrated to 2MASS, are generally in good agreement.  While
larger offsets are seen in a few cases (NGC 6528, NGC 6553, NGC 6638, NGC
6642), the direct
comparison with 2MASS in Fig. \ref{comp2mass} gives no reason to be doubtful
about the calibration of these clusters.  Specifically, 
the mean magnitude offset between
the VVV and 2MASS photometry (weighted by the inverse square of their total
photometric uncertainties) over the magnitude range of stars used for 
calibration is $<$0.016 mag in $J$ and $K_{S}$ and $<$0.023 mag in $H$ 
for these four clusters 
(these mean differences are $<$0.02 mag for 
all other clusters in all bandpasses as well).  

Photometric analysis of GGCs towards the Galactic bulge can be severely 
hampered by contamination from field stars in the bulge and disk, particularly
in cases where bulge and disk contaminants are inseparable from the cluster
evolutionary sequences using colour-magnitude criteria alone.
Statistical field star decontamination methods which compare the
colour-magnitude loci of cluster and field stars generally rely on the 
assumption that reddening is spatially invariant (see Sect.~\ref{decontsect}
below), 
so before undertaking analyses of
the GGC photometry, we first correct for differential reddening and then apply a
statistical field star decontamination procedure. 

\subsection{Differential Reddening}
\label{diffredsect}

We correct our photometric catalog of each cluster for reddening
only in a strictly differential sense (we do not correct for \textit{total}
line of sight extinction).  This is done using the reddening maps 
of \citet{gonzalez}\footnote{The BEAM calculator can be found at
  http://mill.astro.puc.cl/BEAM/calculator.php}, adopting the value of $E(J-K_{S})$ 
corresponding to the location of the cluster center as a reference zeropoint
for the differential reddening corrections over the spatial area of each cluster.  This 
reference value is 
given as $E(J-K_{S})_{REF}$ in Table \ref{tab:cleanparams}.
The photometric catalog for each cluster is then corrected for reddening variations over the field
of view using the \textit{difference} between the value of $E(J-K_{S})$ at a given spatial location
and $E(J-K_{S})_{REF}$ (i.e.~the value at the spatial location of the cluster center).
However, since the \citet{gonzalez} maps were constructed by
measuring the variation
in the $(J-K_{S})$ colour of the Galactic bulge red clump (RC) as a function of
spatial location,   
the number statistics necessary to reliably measure the bulge RC 
colour restrict 
the spatial resolution of the \citet{gonzalez} maps to $>$1 arcmin, while 
significant differential reddening towards bulge GGCs can 
occur on spatial scales of arcseconds 
\citep{javier12,massari,cohen6544}.  Furthermore, the \citet{gonzalez} 
maps were constructed
from aperture photometry catalogs rather than PSF photometry, and therefore
suffer from crowding and incompleteness significantly brightward of their
detection limits as compared to 
PSF photometry \citep[e.g.][see their fig.~6]{mauropipe}.  
Therefore, where available, we have combined the \citet{gonzalez} maps
of the field surrounding each cluster with high spatial resolution
reddening maps (constructed using cluster stars) of the central region of 
the cluster.  
The high resolution maps were taken from
 \citet{javier12} where available (8 clusters), from \citet{cohen6544} in the
 case of NGC 6544, and for 6 more clusters, we employ maps 
similarly constructed from
 archival optical HST imaging described in detail elsewhere (R.~E.~Cohen et
 al., in prep.)\footnote{For comparison, we note that these maps have a median
 spatial resolution of $\sim$10$\arcsec$.}.   
While the high resolution maps are generally restricted to the inner regions of 
the target clusters where the membership probability is high, 
we note that they
extend well beyond the cluster half-light radii from the 
Harris (1996, 2010 revision, hereafter \citealt{h96}) catalog, 
encompassing the majority of cluster members\footnote{The only possible exception is NGC6558, for which the \citet{javier12} map has a radial limit of 1.81$\arcmin$ versus
a half-light radius of 2.15$\arcmin$ from \citet{h96}, although this value 
may not be too reliable as this cluster is core-collapsed \citep{trager}.}.   
These high-resolution maps are also applied in a strictly differential sense, 
relative to $E(J-K_{S})_{REF}$, but
we must take into account that the differential reddening corrections given by the
\citet{javier12} maps may not be 
referred to the same differential reddening zeropoint (i.e.~the cluster center).
Therefore, 
we shift the \citet{javier12} corrections to refer to our reference value of $E(J-K_{S})_{REF}$ (i.e.~the \citealt{gonzalez} 
value at the cluster center) by comparing, for
all stars within the radius permitted by the \citet{javier12} maps, the
$(J-K_{s})$ colour obtained after performing the \citet{javier12} correction
with that resulting from the \citet{gonzalez} correction.  This yields the
mean difference $\Delta$$E(J-K_{S})$ (and standard deviation) 
between the two maps, given in Table \ref{tab:cleanparams}
for clusters in our sample with high resolution maps from \citet{javier12}.  
For the two target clusters with no available high spatial resolution
reddening maps (NGC 6569 and NGC 6638), we employ only the \citet{gonzalez}
maps, noting that they predict quite modest differential reddening over
the entire sampled area 
in both cases ($\Delta$$E(J-K_{S})\leq$0.065).  

\subsection{Field Star Decontamination}
\label{decontbigsect}
\subsubsection{Methodology}
\label{decontsect}
We clean our differential reddening corrected cluster CMDs of field stars 
using a statistical technique detailed in \citet{bonatto1}, 
including recent improvements described by \citet{bonatto3}.  The application
of this technique to VVV PSF photometry is described in 
\citet{cohen6544}, but can be summarized as follows:
Two spatial regions are selected, the first being the spatial region 
to be decontaminated (over which high spatial resolution differential reddening
maps are available) which has area $A_{clus}$ and a total number of stars
$N_{tot}$ in the magnitude range considered for decontamination (see below).
The second area is the comparison (e.~g.~field) region, which has area
$A_{fld}$, which we have chosen to have an inner radius equal to the cluster  
\citealt{h96} tidal radii\footnote{For NGC 6569 and NGC 6638, which lack
high resolution differential reddening maps, we set the cluster area to have
limiting radii of r$\leq$1.90$\arcmin$ and 1.55$\arcmin$ respectively from
the cluster center, corresponding to more than twice the \citet{h96} 
half-light radii in both cases.}.  To statistically decontaminate the cluster
region, the CMD of the cluster region is compared to the CMD of the comparison
region by dividing their CMDs into a
three-dimensional grid of cells in $J,(J-K_{S}),(J-H)$.  The effects of
photometric incompleteness are minimized by including only stars which lie
brightward of the observed cluster area LF peak $J_{lim}$.  
In each CMD cell of the cluster region, the
number of field stars to be removed is calculated by summing the probability
density distributions of all comparison field stars in the analogous CMD cell, 
corrected for
the ratio of cluster to comparison field areas.  This number of
stars, rounded to the nearest integer, are randomly removed from the 
cluster region CMD cell,
and the entire procedure is repeated over $3^6$=729 iterations in which the
cell sizes and locations are varied to mitigate the effects of binning.  
The mean number of surviving cluster stars $N_{clus}$ is calculated over all 
iterations, 
stars are sorted by their survival frequency, and cluster stars are retained 
in order of decreasing survival frequency until this mean number of surviving
cluster stars is reached.  
The efficiency of this field
star decontamination procedure may be gauged using the subtraction efficiency
$f_{sub}$, which is the fraction of (decimal) stars to be subtracted 
(based on the stellar density of the comparison field and the 
ratio of comparison to cluster field area) to the actual (integer) number of 
probable field stars removed from the cluster region.  To attain the highest
possible subtraction efficiencies, the comparison regions generally consist of
an annulus wide enough that $A_{fld}$ is many
times larger than $A_{clus}$.  A large comparison region has the added
advantage that any small-scale variations in the stellar density of the 
comparison field are averaged out, as the comparison regions we employ have
typical areas $\gtrsim$10$^3$ arcmin$^2$.  
However, especially given the relatively
large ($\sim$30$\arcmin$) tidal radii of some of our target clusters, 
in practice an upper limit
to the size of the comparison region is necessary due to several factors.
These include the proximity of other
nearby features not representative of the cluster line of sight such as other
globular and open clusters, and in the case of M69, proximity to the edge of
the VVV survey area over which photometry is available.  The values
of $N_{tot}$, $N_{clus}$, the ratio of comparison to cluster region areas 
$A_{fld}$/$A_{clus}$, the total comparison region area $A_{fld}$, the
subtraction efficiency $f_{sub}$ and the faint magnitude limit $J_{lim}$ are
given for all of our target clusters, including results for NGC 6544 from
\citet{cohen6544} which we add to our sample, 
in Table \ref{tab:cleanparams}, along
with formal uncertainties which take into account both photometric errors and 
Poissonian uncertainties of the total number of stars in the cluster
and comparison regions.  
The impact of uncertainties in the decontamination procedure on the photometric
features which we measure are discussed in the context of each of these
features in Sects.~\ref{tipsect}, \ref{bumperrsect} and \ref{slopeerrsect}.   
 
\begin{table*}
\caption{Differential Reddening and Decontamination Parameters}
\begin{tabular}{rcccccccccc}
\hline
Cluster & $E(J-K_{S})_{REF}$ & $\langle\Delta$$E(J-K_{S})\rangle$$^{a}$ &
Reddening Map$^{b}$ & $N_{tot}$ & $N_{clus}$ & A$_{fld}$/A$_{clus}$ &
A$_{fld}$ &  $f_{sub}$ & $J_{lim}$ \\
 & mag & mag & & & & & arcmin$^2$ & \% & mag \\
\hline
NGC6380 & 0.496 & & 1 & 5169$\pm$72 & 4292$\pm$148 & 117.67 & 854.10 & 99.8$\pm$0.3 & 18.40 \\
NGC6401 & 0.417 & & 1 & 11385$\pm$107 & 6086$\pm$403 & 61.95 & 732.46 & 99.3$\pm$0.4 & 18.40 \\
NGC6440 & 0.530 & & 1 & 6413$\pm$80 & 4443$\pm$164 & 49.77 & 366.00 & 99.8$\pm$0.1 & 18.40 \\
NGC6441 & 0.205 & & 1 & 15050$\pm$123 & 10860$\pm$120 & 39.07 & 495.92 & 99.9$\pm$0.1 & 18.80 \\
NGC6453 & 0.285 & & 1 & 6616$\pm$81 & 4717$\pm$46 & 149.67 & 1100.67 & 99.9$\pm$0.1 & 18.20 \\
NGC6522 & 0.234 & -0.018$\pm$0.017 & 2 & 21244$\pm$146 & 14942$\pm$591 & 75.13 & 1281.29 & 99.9$\pm$0.1 & 18.40 \\
NGC6528 & 0.271 & & 1 & 11857$\pm$109 & 4231$\pm$209 & 32.74 & 367.41 & 99.7$\pm$0.1 & 18.40 \\
NGC6544 & 0.736 & & 3 & 24166$\pm$155 & 7771$\pm$235 & 32.80 & 368.57 & 93.7$\pm$1.2 & 18.84 \\
NGC6553 & 0.369 & 0.008$\pm$0.029 & 2 & 48836$\pm$221 & 29283$\pm$412 & 11.63 & 558.43 & 99.8$\pm$0.1 & 18.20 \\
NGC6558 & 0.150 & 0.004$\pm$0.016 & 2 & 6678$\pm$82 & 3707$\pm$87 & 87.20 & 897.46 & 99.5$\pm$0.2 & 18.60 \\
NGC6569 & 0.199 & & 4 & 7692$\pm$88 & 5744$\pm$33 & 32.35 & 366.92 & 99.9$\pm$0.1 & 18.60 \\
NGC6624 & 0.104 & 0.003$\pm$0.021 & 2 & 24038$\pm$155 & 14537$\pm$574 & 26.57 & 1282.67 & 99.9$\pm$0.1 & 19.20 \\
M28 & 0.138 & 0.014$\pm$0.027 & 2 & 56829$\pm$238 & 33320$\pm$88 & 16.56 & 1274.84 & 99.9$\pm$0.1 & 18.60 \\
M69 & 0.017 & 0.004$\pm$0.009 & 2 & 14107$\pm$119 & 5079$\pm$389 & 0.53 & 17.61 & 98.1$\pm$0.6 & 18.68 \\
NGC6638 & 0.189 & & 4 & 9654$\pm$98 & 3466$\pm$358 & 4.66 & 35.19 & 99.1$\pm$0.2 & 18.93 \\
NGC6642 & 0.161 & 0.002$\pm$0.027 & 2 & 9853$\pm$99 & 6626$\pm$76 & 39.79 & 616.03 & 99.9$\pm$0.1 & 19.40 \\
M22 & 0.000 & 0.042$\pm$0.044 & 2 & 153216$\pm$391 & 65122$\pm$4052 & 1.69 & 224.28 & 96.5$\pm$0.8 & 18.45 \\
\hline
\multicolumn{11}{l}{$^{a}$ Reddening map zero point offset in the sense \citep{javier12}-\citep{gonzalez}}\\
\multicolumn{11}{l}{$^{b}$ Reddening maps applied to cluster photometry before
  decontamination as follows: (1) Cohen et al.~2016 (in prep.) (2)\citealt{javier12}} \\
\multicolumn{11}{l}{(3)\citealt{cohen6544} (4)\citealt{gonzalez}}\\
\end{tabular}
\label{tab:cleanparams}
\end{table*}

\subsubsection{Proper Motions: An Independent Test of the Decontamination
  Algorithm}

\begin{figure}
\includegraphics[width=0.5\textwidth]{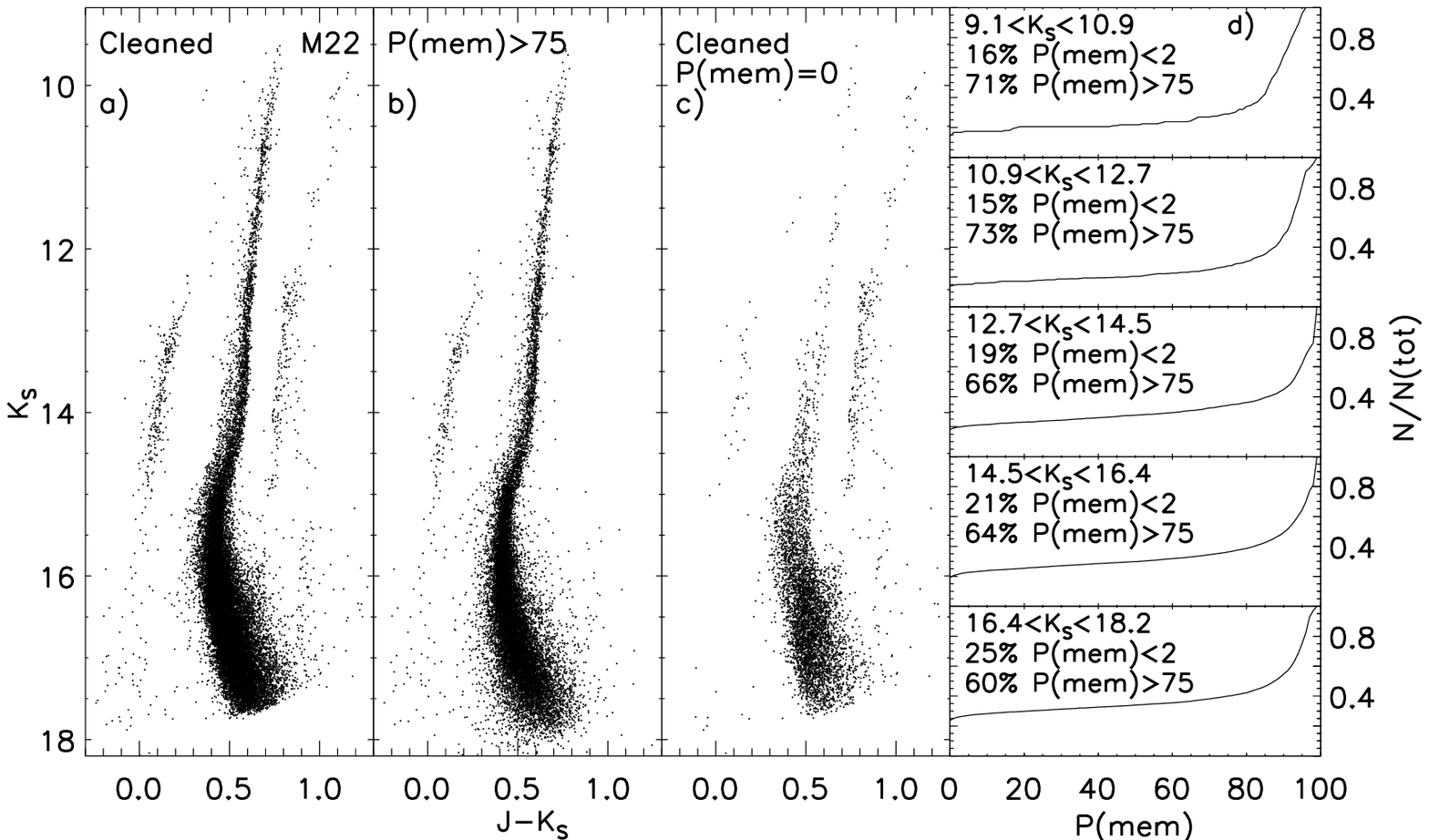}
\caption{\textbf{(a:)} CMD of all stars present in the proper motion
catalog of \citet{libralato} which passed the statistical decontamination
procedure described in Sect.~\ref{decontsect}.  \textbf{(b:)} All stars in our
photometric catalog which are likely proper motion members ($P_{mem}$$>$75) 
according to \citet{libralato}. \textbf{(c:)} Stars which survived our
statistical decontamination algorithm but have a proper motion based
membership probability of 0 from \citet{libralato}. \textbf{(d:)} Cumulative distribution
of \citet{libralato} membership probability for all stars which survived
the statistical decontamination procedure, shown in five magnitude bins.}
\label{m22pmfig}
\end{figure}

As an independent test of the decontamination procedure, we may compare
our statistically decontaminated CMDs with results from relative proper motion
studies. 
There is one cluster in our sample, M 22,   
for which membership probabilities have been calculated from relative proper
motions over a relatively wide field of view by
\citet{libralato}\footnote{Proper motions in M22 were also published by
  \citet{m22pms}, although they did not give formal membership probabilities.}.
After matching our photometric catalog to theirs, in Fig.~\ref{m22pmfig} 
we compare all stars in our
(differential reddening corrected) catalog surviving statistical
decontamination (shown in panel a) with those which
\citet{libralato} considered likely members (panel b) as well as those
which survived the decontamination procedure but have zero probability of
membership according to their proper motions (panel c).  
It is evident that for this cluster, 
the decontamination procedure fails to remove a minority of
field RGB stars, seen 0.2-0.3 mag redward of the cluster RGB 
in panels (a) and (c) of Fig.~\ref{m22pmfig}.  There are several probable
causes for this effect (see below), and proper motion selection
can be similarly subject to contamination from field stars with cluster-like
proper motions \citep[e.g.][]{libralatovvv}, although it may be possible to 
take this effect into account statistically in some cases 
\citep[e.g.][]{milonebinfrac}.

To further compare the performance of the
decontamination algorithm versus the use of proper motions as a function of
magnitude (or, equivalently, photometric error), in panel (d) of
Fig.~\ref{m22pmfig} we divide the stars in our catalog which survived the
decontamination algorithm into magnitude bins.  In each magnitude bin, we plot
the cumulative distribution of the proper motion membership probabilities
from \citet{libralato}, as well as giving the 
fraction of surviving stars in each bin which fall into the ranges of proper
motion probability used by \citet{libralato} to identify definite members
($P_{mem}$$>$75\%) and definite non-members ($P_{mem}$$<$2\%).  It is clear from
the right-hand panel of Fig.~\ref{m22pmfig} that
the contamination rate among our statistically decontaminated
sample is $\leq$25\% \textit{without} the use of a colour cut, and  
this contamination rate does not vary appreciably with magnitude. 

\subsection{Colour-Magnitude Diagrams}
 
CMDs of all of our target clusters are shown in the $(K_{S},J-K_{S})$ plane in Fig.~\ref{cmdsjk}
and in the $(J,J-H)$ plane in Fig.~\ref{cmdsjh}, and are also included
  in the supplementary figures along with the RGB LFs.  Stars which passed the
decontamination procedure are shown in black, whereas stars which failed are
shown in grey.  In addition, we have identified known variables in our target
clusters by matching our 
2MASS-astrometrized $JHK_{S}$ catalogs 
with the most recent version 
of the Catalog of Variable Stars in Galactic Globular Clusters
\citep{clement}\footnote{http://www.astro.utoronto.ca/$\tilde{
  }$cclement/read.html} and the catalog of equatorial coordinates 
by \citet{samus}.  These variables are overplotted on the CMDs as blue
diamonds.  We have excluded known variables from the determination of the
fiducial sequences since detailed variability studies show that AGB variables
may be present faintward of the RGB tip, and a more thorough
discussion of variability on the upper RGB and the inclusion of variables in
RGB tip magnitude measurements can be found in Sect.~\ref{tipsect}).  
In any case, the influence of known variables on each of the 
photometric features we measure is discussed in the context of each of the relevant features in Sect.~\ref{anasect}.    

While in a minority of cases the decontamination procedure results 
in gaps in the cluster
evolutionary sequences or a failure to remove field RGB stars, this is a 
likely consequence of the differing spatial resolution
between the reddening maps applied to the comparison field ($>1$$\arcmin$;
\citealt{gonzalez}) and those applied to the cluster regions before
decontamination (see \citealt{javier12},\citealt{cohen6544},
Sect.~\ref{diffredsect}).  In addition, the fact that the clusters which are
most susceptible to this effect (NGC 6553 and M22) are also the most nearby
along the line of sight suggests that this could also be partially due to
preferential obscuration of the field (e.g.~bulge) population by the cluster
in these cases, and we note that the \citet{gonzalez} map which we employ for 
the comparison field tends to overestimate reddening at small ($<$4 kpc) 
heliocentric distances \citep{matthiasgonzalez}.  In any case, we find that
the location of photometric features we measure is insensitive to this effect 
beyond their reported uncertainties, 
based both on comparisons to previous studies employing radial
cuts and/or proper motions (see Sect.~\ref{comparsect}) as well as a
comparison between values measured using statistically decontaminated CMDs 
versus field-subtracted LFs (see Sect.~\ref{bumperrsect}).

\begin{figure*}
\includegraphics[width=0.98\textwidth]{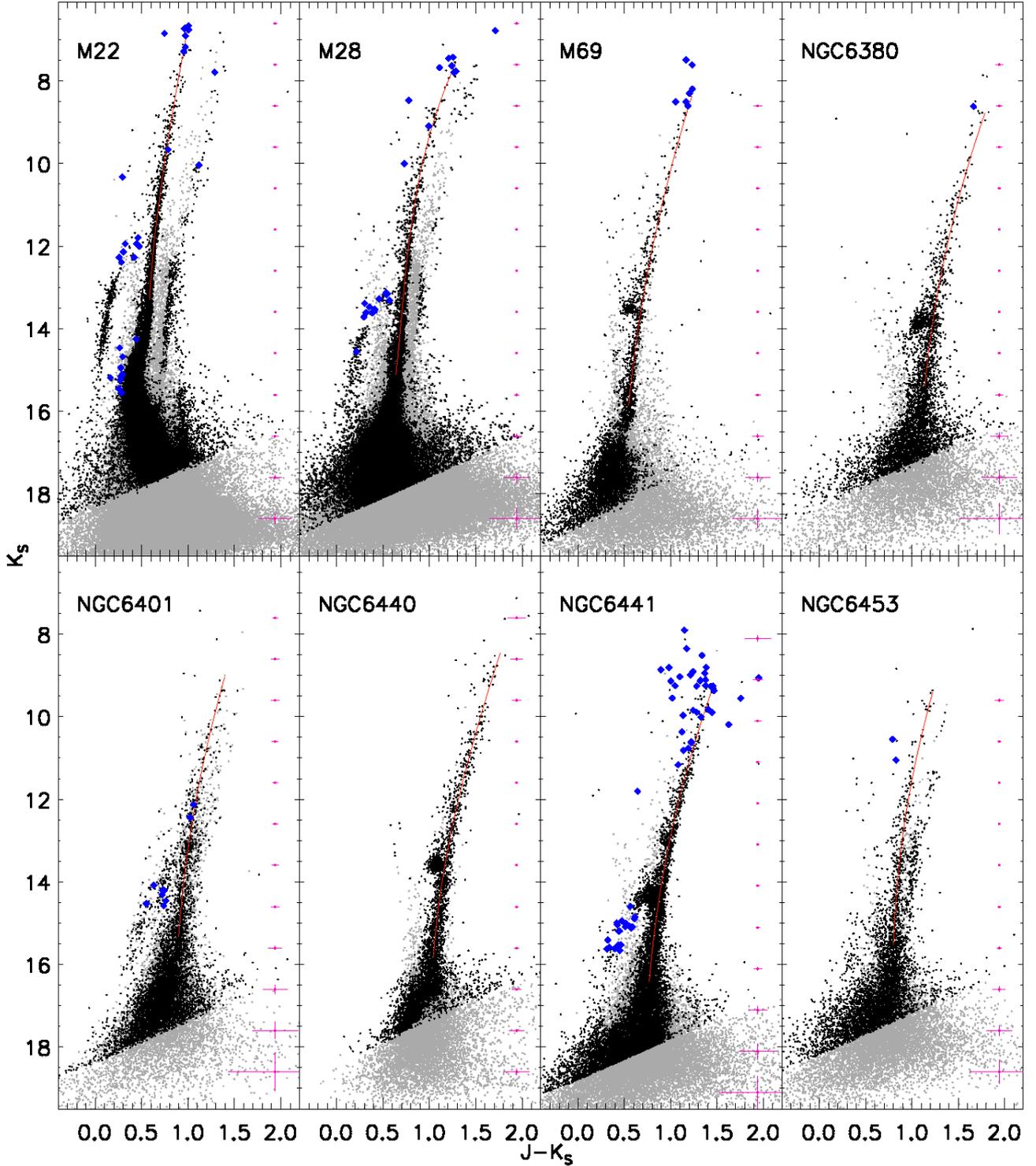}
\caption{Differential reddening corrected CMDs of all of our target clusters 
in the $(K_{S},J-K_{S})$ plane.
  All stars within the cluster region are shown in grey, those surviving the
  decontamination procedure are shown in black, and known variables which were
  removed from our analysis are plotted as blue diamonds.  
  In addition, our fiducial sequences 
  are shown in red, and median photometric errors in magnitude bins are 
  shown along the right-hand side of each CMD in magenta.  High-quality
  $(K_{S},J-K_{S})$ and $(J,J-H)$ CMDs for all target clusters are included in
as a set of supplementary figures.}
\label{cmdsjk}
\end{figure*}
\begin{figure*}
\includegraphics[width=0.98\textwidth]{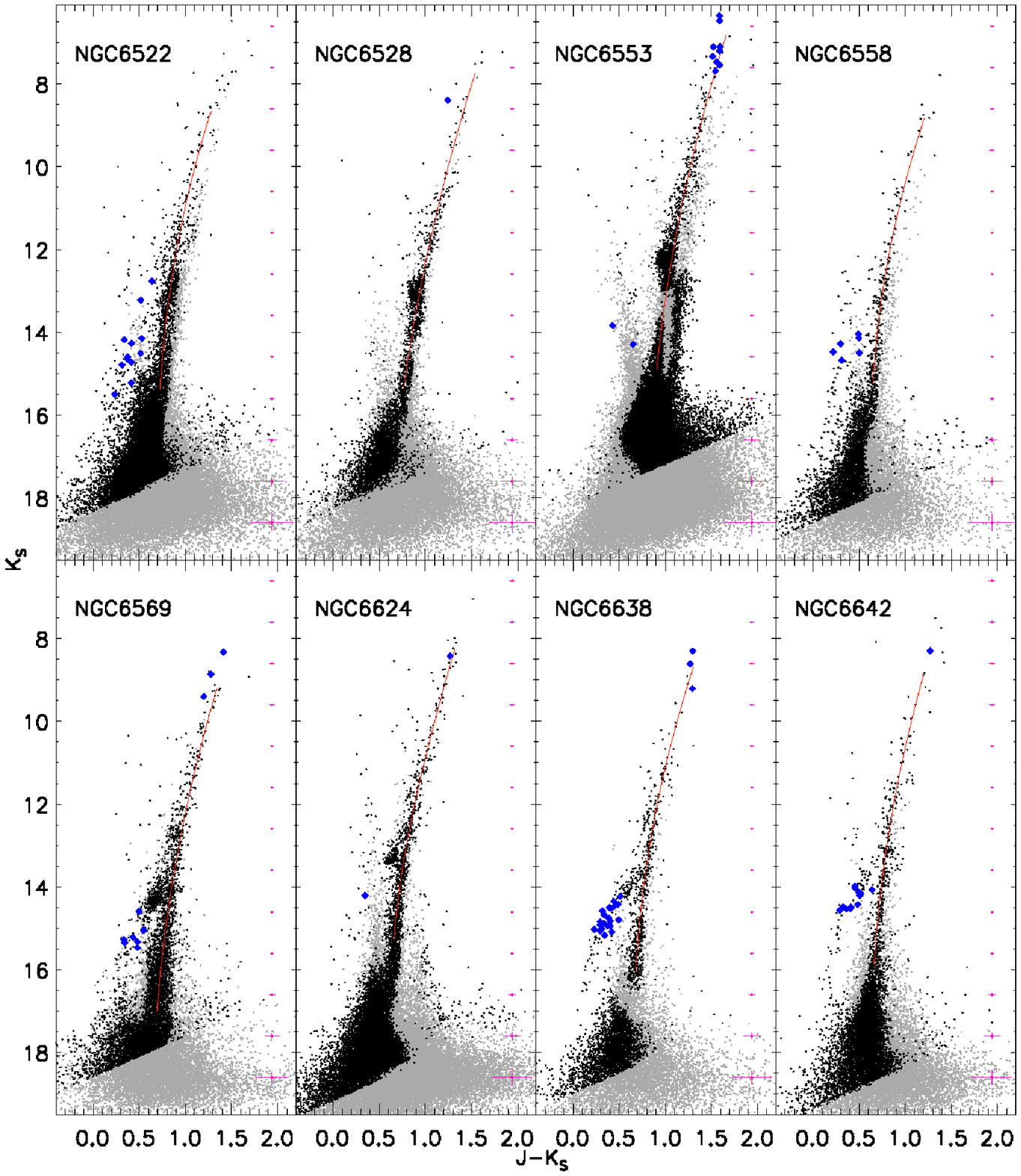}
\contcaption{}
\end{figure*}

\begin{figure*}
\includegraphics[width=0.98\textwidth]{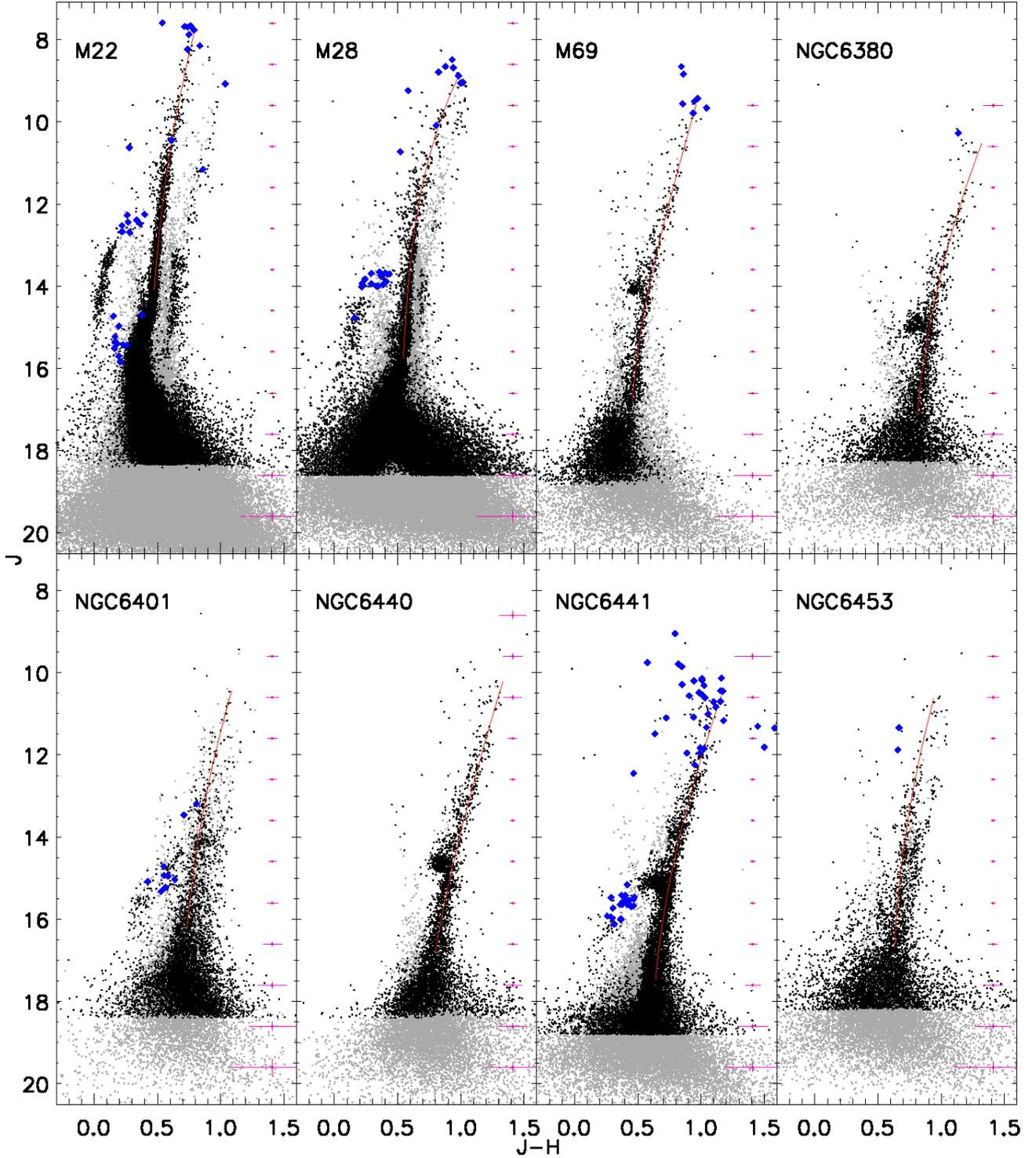}
\caption{As for Fig.~\ref{cmdsjk} but in the $(J,J-H)$ plane.}
\label{cmdsjh}
\end{figure*}

\begin{figure*}
\includegraphics[width=0.98\textwidth]{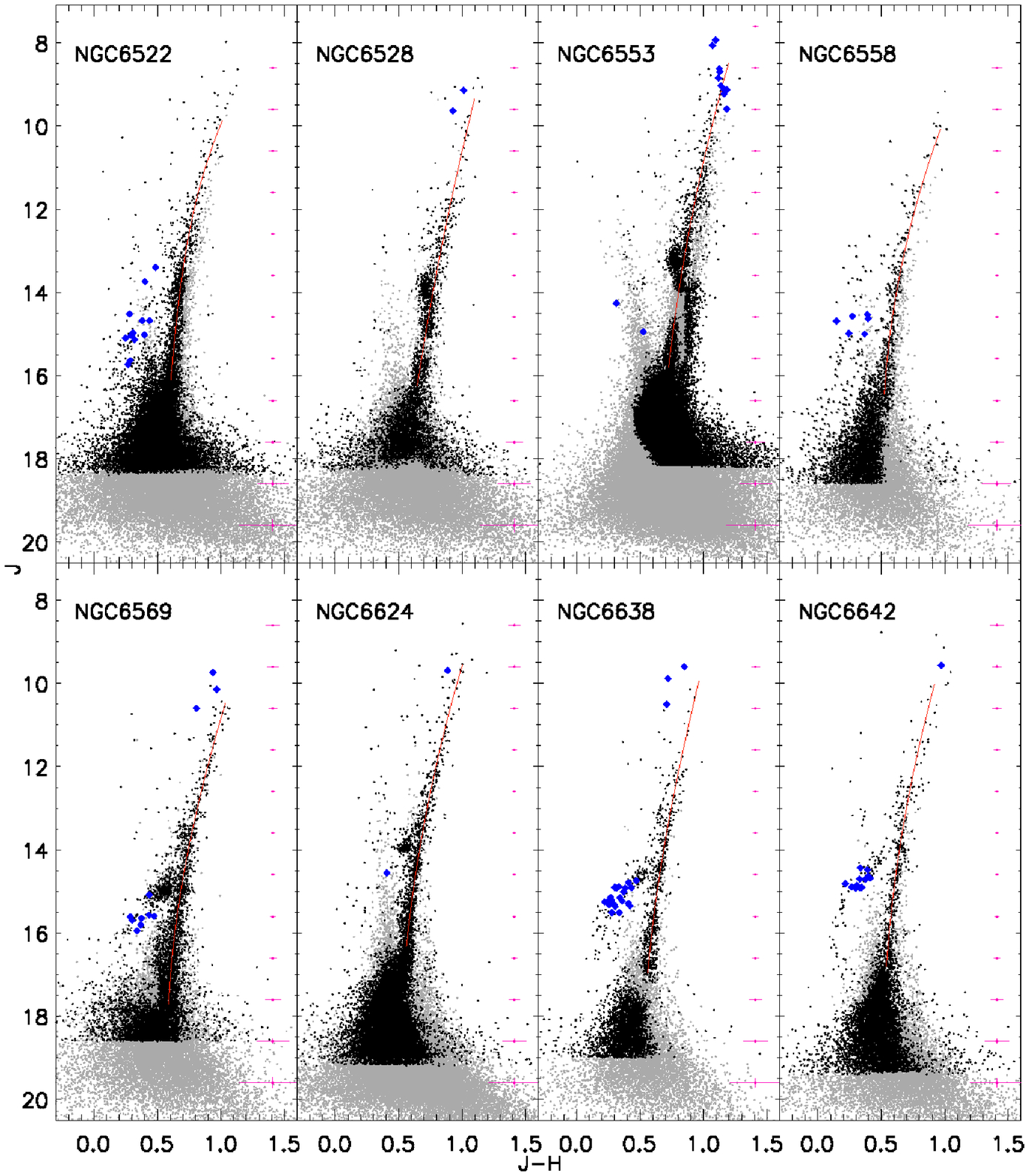}
\contcaption{}
\end{figure*}

\section{Observed Photometric Features}
\label{anasect}

\subsection{Cluster Fiducial Sequences}

In order to derive calibrations between cluster chemical abundances and
photometric features along the cluster RGBs, 
including the RGB tip, bump and slope, we
fit fiducial sequences to the RGBs in the differential reddening corrected,
field star decomtaminated CMDs, which are hereafter referred to as the
``processed'' CMDs.  Fiducial
sequences are fit
using an iterative procedure similar to previous studies \citep{ferraro00,v04obs,cohen6544,cohenispi}.  
First, a rough visual 
colour-magnitude cut was used to isolate the CMD region of the RGB.  
Next, the RGB was divided into magnitude bins of width 0.5 mag, and the 
median colour and magnitude in each bin was measured.  A low
order ($\leq$3) polynomial was then fit to these median colours as a function of
magnitude, iteratively rejecting stars more than 2$\sigma$ in colour from the 
fit polynomial in each bin.
This process is repeated until convergence is indicated by the number of
surviving stars changing by under 2\% since the previous iteration.   
This procedure is still necessary even if the field star decontamination
algorithm functions perfectly, since bona fide cluster HB and AGB stars should
still be present and thus can be removed from consideration in a statistical 
manner to construct sequences representative of the RGB.

\begin{figure}
\includegraphics[width=0.49\textwidth]{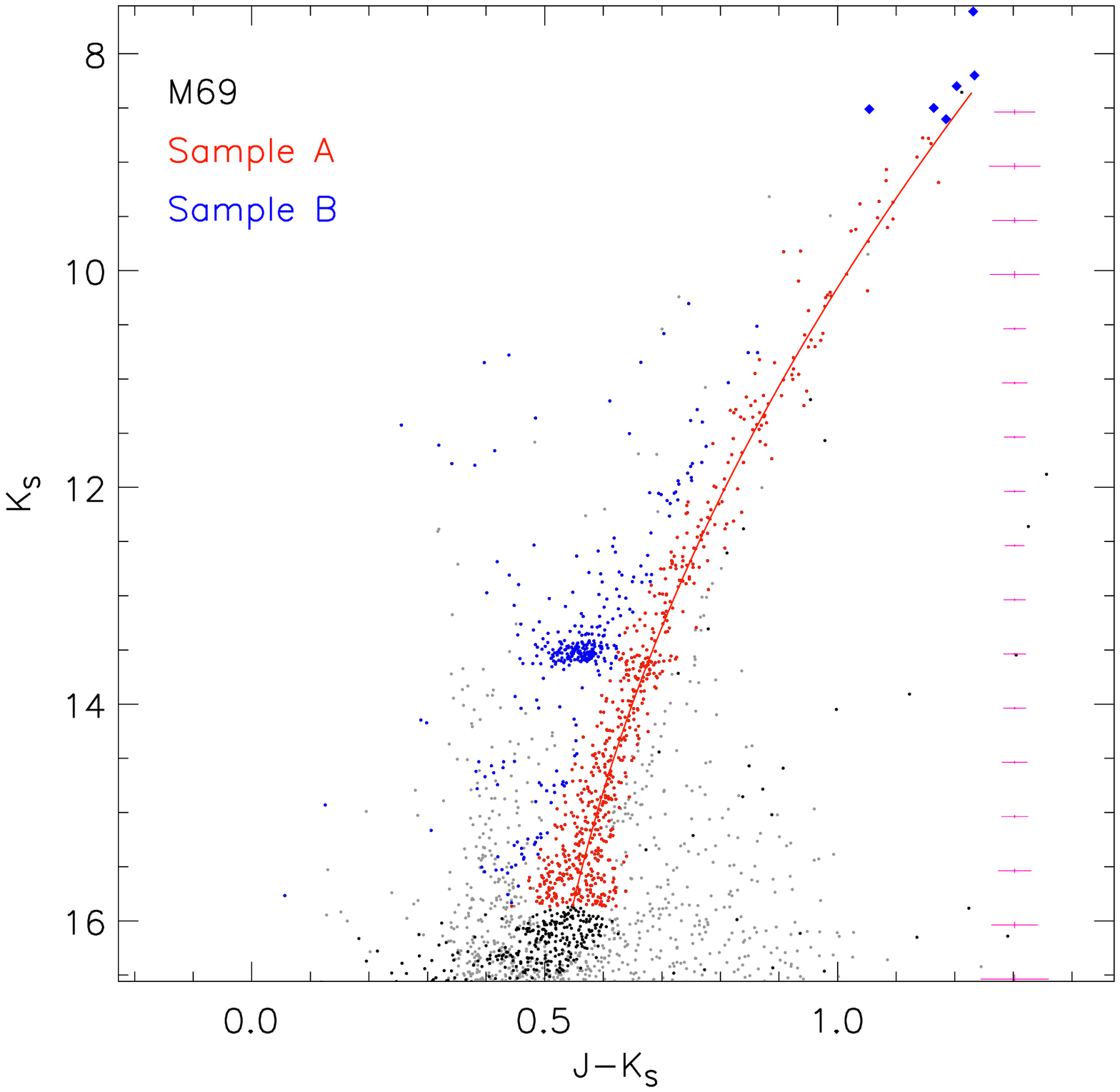}
\caption{The processed CMD of M69, illustrating the selection of stellar
  sub-samples used to measure CMD features.  
  Sample A, used to measure the RGB slope and the RGBB magnitude, is shown in
  red, and Sample B, used to measure the magnitude of the HB, is shown in
  blue.  Other symbols are as in Fig.~\ref{cmdsjk}.}
\label{fidexample}
\end{figure}

Once the fiducial sequence has been constructed, we make a colour cut
  in the $(K_{S},J-K_{S})$ CMD to
identify subsamples of stars used to measure the slope of the upper RGB as
well as the locations of the red giant
branch bump (RGBB) and the horizontal branch (HB).  Specifically, in order to
minimize contamination of the cluster RGB by the HB,  
we use only stars with
$(J-K_{S})$ colours within 3$\sigma$ of the fiducial sequence 
(where $\sigma$ represents the median photometric error as a function of
$K_{S}$ magnitude), which we refer to as Sample A.  Similarly, to avoid a bias
on the HB magnitude caused by the RGBB, we measure the location of the HB
using only stars blueward of Sample A, which we refer to as Sample B.  An
example of the selection of both of these samples is shown for the case of M
69 in Fig.~\ref{fidexample}.

The coefficients of the fiducial sequences in both the $(K_{S},J-K_{S})$ and
$(J,J-H)$ CMDs are given in Table \ref{fidcoefftab} along with their ranges of
validity.  We reiterate that these fiducial sequences are derived from photometry which
has been corrected for differential reddening across the cluster relative to the 
cluster center, but has not been corrected for \textit{total} line-of-sight extinction.

\begin{table*}
\caption{Coefficients of Observed Fiducial Sequences}
\begin{tabular}{rcccccc}
\hline
\multicolumn{7}{c}{$(J-K_{S})=a_{0}+a_{1}K_{S}+a_{2}K_{S}^{2}+a_{3}K_{S}^{3}$} \\
\hline
Cluster & $K_{S,min}$ & $K_{S,max}$ & $a_{0}$ & $a_{1}$ & $a_{2}$ & $a_{3}$ \\
\hline
 NGC6380 &   8.770 &  15.609 &    3.7852440 &   -0.3019084 &    0.0084423 &    0.0000000 \\
 NGC6401 &   9.040 &  15.728 &    3.1537068 &   -0.2626610 &    0.0075405 &    0.0000000 \\
 NGC6440 &   8.459 &  15.986 &    3.0384719 &   -0.1304930 &   -0.0054349 &    0.0003610 \\
 NGC6441 &   9.188 &  16.612 &    3.6435592 &   -0.3149566 &    0.0085259 &    0.0000000 \\
 NGC6453 &   9.402 &  15.720 &    3.1904884 &   -0.2939827 &    0.0090236 &    0.0000000 \\
 NGC6522 &   8.649 &  15.589 &    3.4198779 &   -0.3620996 &    0.0147217 &   -0.0001667 \\
 NGC6528 &   7.739 &  15.560 &    2.9406701 &   -0.2207616 &    0.0051508 &    0.0000000 \\
 NGC6553 &   6.812 &  15.106 &    2.8970294 &   -0.2216789 &    0.0059164 &    0.0000000 \\
 NGC6558 &   8.822 &  15.551 &    3.2531687 &   -0.3158048 &    0.0095168 &    0.0000000 \\
 NGC6569 &   9.192 &  17.227 &    3.0955093 &   -0.2485383 &    0.0063109 &    0.0000000 \\
 NGC6624 &   8.234 &  15.478 &    2.2191760 &   -0.0562974 &   -0.0103873 &    0.0004842 \\
     M28 &   7.548 &  15.314 &    4.9231014 &   -0.8379519 &    0.0577563 &   -0.0013936 \\
     M69 &   8.358 &  16.068 &    2.8374305 &   -0.2459917 &    0.0064082 &    0.0000000 \\
 NGC6638 &   8.682 &  16.256 &    3.1853596 &   -0.2878985 &    0.0082148 &    0.0000000 \\
 NGC6642 &   8.830 &  17.089 &    2.8574505 &   -0.2486005 &    0.0069500 &    0.0000000 \\
     M22 &   6.722 &  13.500 &    1.5011816 &   -0.0506029 &   -0.0058926 &    0.0003390 \\
\hline
\multicolumn{7}{c}{$(J-H)=a_{0}+a_{1}J+a_{2}J^{2}+a_{3}J^{3}$} \\
\hline
Cluster & $J_{min}$ & $J_{max}$ & $a_{0}$ & $a_{1}$ & $a_{2}$ & $a_{3}$ \\
\hline
 NGC6380 &  10.526 &  17.123 &    3.4250948 &   -0.2755087 &    0.0071731 &    0.0000000 \\
 NGC6401 &  10.436 &  16.290 &    2.6685162 &   -0.2095136 &    0.0055676 &    0.0000000 \\
 NGC6440 &  10.214 &  16.779 &    2.1937630 &   -0.0479402 &   -0.0058459 &    0.0002235 \\
 NGC6441 &  10.532 &  17.612 &    2.3822459 &   -0.0921838 &   -0.0049885 &    0.0002610 \\
 NGC6453 &  10.622 &  16.802 &    2.4814560 &   -0.2045837 &    0.0056062 &    0.0000000 \\
 NGC6522 &   9.863 &  16.009 &    4.9753892 &   -0.7589820 &    0.0455941 &   -0.0009510 \\
 NGC6528 &   9.338 &  16.265 &    2.0592060 &   -0.1242321 &    0.0022911 &    0.0000000 \\
 NGC6553 &   8.499 &  15.715 &    2.2585166 &   -0.1573739 &    0.0038187 &    0.0000000 \\
 NGC6558 &  10.079 &  16.537 &    2.9453618 &   -0.2744620 &    0.0077358 &    0.0000000 \\
 NGC6569 &  10.471 &  17.562 &    2.1521419 &   -0.0924727 &   -0.0036560 &    0.0002197 \\
 NGC6624 &   9.536 &  16.365 &    2.3521783 &   -0.1859118 &    0.0046656 &    0.0000000 \\
     M28 &   8.825 &  15.205 &    4.4994024 &   -0.7116985 &    0.0437171 &   -0.0009187 \\
     M69 &   9.574 &  16.907 &    1.3614957 &    0.0510122 &   -0.0142585 &    0.0004771 \\
 NGC6638 &   9.936 &  16.999 &    1.7881424 &   -0.0704574 &   -0.0028577 &    0.0001619 \\
 NGC6642 &  10.023 &  17.402 &    2.2953396 &   -0.1858459 &    0.0048425 &    0.0000000 \\
     M22 &   7.737 &  14.092 &    1.6818249 &   -0.1488596 &    0.0044827 &    0.0000000 \\
\hline
\end{tabular}
\label{fidcoefftab}
\end{table*}

\subsection{Red Giant Branch Tip (TRGB)}
\label{tipsect}
For the clusters in common with \citet{v10}, we could in principle use their
measured TRGB magnitudes since both studies are similarly reliant upon
2MASS photometry at these bright magnitudes, and they likewise applied a
statistical procedure to remove field stars from 2MASS photometry.  
However, we redetermine these magnitudes for three reasons.  First, application
of differential reddening corrections could change these values somewhat
(although this effect would likely be small due to the horizontality of the
reddening vector in $K_{S}$,$(J-K_{S})$).   
Second, we ensure that all
clusters (not just those in common with \citealt{v10}) have their TRGB
magnitudes measured self-consistently.  Third, we avoid luminous AGB 
variables unknown in previous investigations (but see below).    
Therefore, when identifying the location of the TRGB, we use the
processed CMDs as they take 
the photometry in all three $JHK_{S}$ bands 
as well as photometric errors into consideration, 
and select the brightest star along the cluster RGB in both the
$K_{S},(J-K_{S})$ and $J,(J-H)$ decontaminated CMDs.  In addition, we have
checked the candidate TRGB star in each cluster against unmatched stars from
2MASS within the \citet{h96} tidal radii as well as unmatched stars in the 
catalogs of
\citet{v10} where available.  

\begin{figure}
\includegraphics[width=0.49\textwidth]{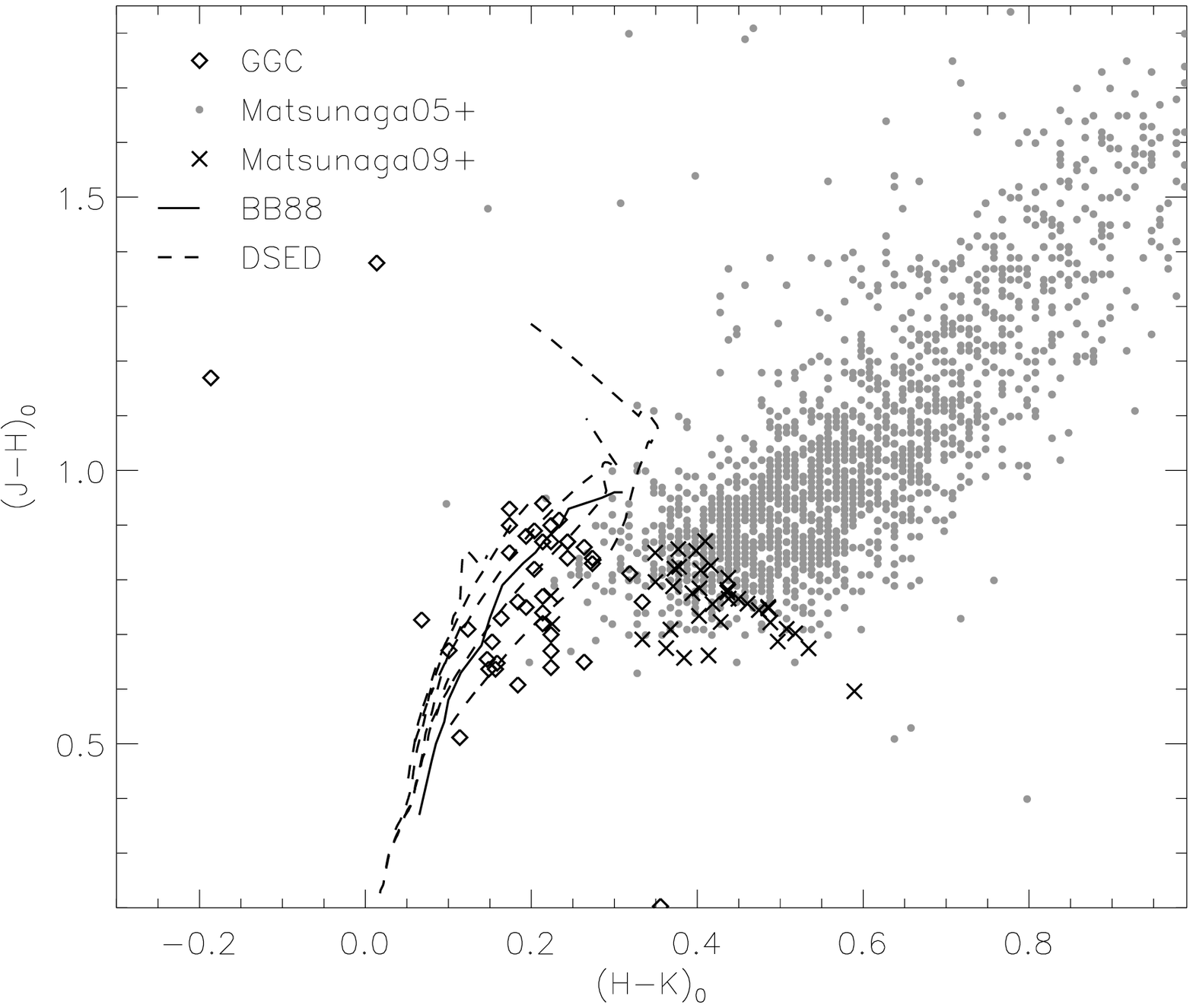}
\caption{Near-IR colour-colour diagram showing dereddened colours of periodic AGB
  variables in several GGCs from \citet{agb362} and \citet{agbm22} as
  diamonds, as well as
  Mira variables towards the Galactic centre from 
  \citet{matsunaga05,matsunaga09} as filled grey circles and
  crosses respectively.  K and M giant colours from \citet{bb88} are
  overplotted as a solid line, and predictions of 12 Gyr DSED models
for $\mathit{[Fe/H]}$=$-$2.5 to +0.5 (increasing from left to right on the plot) 
as dashed lines.}
\label{tcd}
\end{figure}

In many cases, selection of the brightest non-AGB cluster member is ambiguous,
so the selected TRGB star  
is ultimately only the \textit{candidate} brightest RGB star.  
Although a statistical uncertainty on the
TRGB magnitude can be estimated based on evolutionary
considerations \citep[e.g.][]{ferraro00}, 
the true uncertainty in the TRGB location may be difficult to 
ascertain for three reasons.
First, the exponential nature of the RGB LF implies that
the RGB is sparsely populated close to the tip.  Even globular clusters
typically have too few RGB stars to employ 
statistical methods for quantifying the
TRGB uncertainty, 
such as the edge detection technique pioneered by \citet{edgedetect} or
maximum likelihood methods (\citealt{trgbml1,trgbml2} but see
\citealt{trgbbayesian}).  Second, it is difficult to separate RGB and AGB
members based on photometry since AGB stars are effectively colocated with the
TRGB in near-IR colour-magnitude \textit{and} colour-colour planes.   
This is illustrated in a near-IR two colour diagram in Fig.~\ref{tcd}, where we plot AGB variables in NGC 362, NGC
2808 and M 22 from \citet{agb362} and \citet{agbm22} as 
diamonds.  We 
have included only periodic variables which those studies do not suspect of
being nonmembers, and observed colours were converted to the dereddened plane 
using
$E(B-V)$ values from \citet{v13} (or \citealt{monaco} in the case of M22) and
the $R_{V}$=3.1 extinction law from Appendix B of \citet{hendricks}.  
To illustrate the coincidence of these variables with GGC RGBs, we overplot 
K and M giant colours from
\citet{bb88} as well as predictions of 12 Gyr $\alpha$-enhanced
($[\alpha/Fe]$=+0.4 for $\mathit{[Fe/H]}$$<$0, otherwise $[\alpha/Fe]$=+0.2)
isochrones over a wide range
of cluster metallicities (-2.5$<[Fe/H]<$+0.5) 
from the Dartmouth Stellar
Evolution Database \citep[DSED][]{dotter08} as these models
reproduce GGC RGB near-IR colours to $\sim$0.03 mag \citep{cohenispi}.
Additionally, dereddened colours of Mira variables towards the Galactic bulge
from the 
surveys of \citet{matsunaga05,matsunaga09} are shown as filled grey circles
and crosses respectively.  Fig.~\ref{tcd} illustrates that 
variability is common close to the TRGB in the near-IR colour-colour plane as
well as optical and near-IR CMDs.  The problem of disentangling bright AGB and 
RGB cluster members 
is not restricted to more metal-rich GGCs, as periodic 
variables likely to be cluster members have also been detected in
GGCs as metal-poor as M 15 \citep{agbm15}, which has $\mathit{[Fe/H]}$=-2.33
\citep{c09}, in addition to the metal-intermediate to
metal-poor GGCs with variables shown in Fig.~\ref{tcd}.

Third,
both AGB and RGB stars are often photometrically
variable, so that even when an AGB star has a colour and/or magnitude which
is separable from the RGB \textit{in the mean}, it may coincide with the RGB
at some pulsational phases.
This is illustrated, for example, in fig.~6 of 
\citet{montegriffo47tuc}, where the location of AGB variables near the RGB tip 
on the CMD changes significantly as a function of their pulsational phase.
However, not only do AGB \textit{and} RGB stars both vary, but it is unclear
whether their pulsational properties can be used to disentangle their
evolutionary state.  For example, the optical variability study of 
47 Tuc by \citet{agb47tuc} found that \textit{all} cluster giants 
which they detected with $(V-I)>$1.8 are variable (see their fig.~3).  
Furthermore, while
upper RGB and AGB variables may be more easily detected owing to generally
larger pulsational amplitudes \citep{kissbeddinglmc,kissbeddingsmc}, lower
amplitude RGB stars pulsate as well, with amplitudes ranging from hundreths of
mags for the OGLE Small
Amplitude Red Giants in the Magellanic Clouds \citep[OSARGs;][]{osarg} and the
Galactic bar \citep{osargbar} down to millimagnitudes for low-luminosity RGB
stars \citep[e.g.][]{beddingsolar}.

Disconcertingly, the success of recent variability campaigns targeted at luminous  
GGC members at both optical \citep[e.g.][]{layden1851,agbm22,abbas6496} and
infrared wavelengths \citep{matsunaga06,sloanmira} 
implies that the
current census of variable upper RGB/AGB stars in GGCs 
is likely incomplete, as these stars are often saturated in photometric
time series investigations of less luminous RR Lyrae and SX Phoenicis pulsators.  
Moreover,  
even when variability data are available, it remains unclear to what extent
pulsational properties aid in separating AGB from RGB members near the TRGB, especially when
only a small number of time-series epochs are available. 
On one hand, fig.~4 of \citet{agb47tuc} as well as the results of
\citet{agbm22} suggest that variability
amplitude decreases with decreasing luminosity, although with a relatively
small sample size and sparse time sampling, the evolutionary state of any
individual case may still be unclear.  Perhaps the most useful link
  between pulsational properties and evolutionary state for luminous giants
  was illustrated using a combination of photometry and extensive time series
  data.  
  To this end, \citep{kissbeddinglmc,kissbeddingsmc} found that 
a significant
fraction of the variables below the TRGB are RGB rather than AGB stars.
In addition, \citet{osarg} managed to efficiently separate RGB and AGB stars
below the TRGB using detailed pulsational properties, revealing that RGB
(e.g.~non-AGB) 
pulsators tend to have almost exclusively short primary periods (P$\lesssim$60d;
see their fig.~8) and small
pulsational amplitudes ($A_{I}$$<$0.14 mag).  
On the other hand, \citet{pulsamp47tuc}
discuss the difficulty of separating AGB and RGB stars using pulsational
properties.  While small-amplitude variables below the TRGB
appear to be dominated by RGB stars \textit{in a statistical
  sense}, an AGB status may be difficult to exclude on any 
individual case-by-case basis, at least when very high-quality time series 
data are lacking.  

The time series aspect of
VVV imaging unfortunately cannot provide any clues with respect to our target
cluster TRGBs, as our VVV PSF photometry saturates $>$1 magnitude 
below the TRGB. 
Therefore,
given the complexities associated with choosing a single star to
represent the location of the TRGB, we provide a detailed cluster-by-cluster 
description of our choice of TRGB star in Appendix \ref{tipdetailsect}, and
list the corresponding TRGB magnitudes in Table \ref{tipstartab} 
without formal uncertainties.  
A comparison between our TRGB magnitudes and those reported in the literature 
is given in Sect.~\ref{comparsect}, and
we discuss empirical constraints on the precision of TRGB measurements 
in Sect.~\ref{bumptipsect}, and the impact of the TRGB uncertainty on RGB
slope measurements in Sect.~\ref{slopesect}.
Lastly, one possibility 
for definitively separating bright RGB and AGB
members in the absence of high-quality time series data
may be via spectroscopy \citep{agbrgbhalpha}.

\subsection{The Red Giant Branch Bump}

The RGB bump (RGBB) in the RGB LF was 
originally described by \citet{iben} and \citet{thomas}, and the investigation
of \citet{fusipecci}
was one of the earlier studies to quantify the relationship between the RGBB 
luminosity and the chemical abundances of cluster stars.  Empirical 
relations between
cluster metallicity and the RGBB luminosity have been 
presented in optical \citep[e.g.][]{natafbump} as well as
near-IR \citep{chobump,v04abs} bandpasses, and we measure the
location of the bump in all three $JHK_{S}$ filters.  While we defer a
discussion of the bump luminosity (and consequently of the GGC
distance scale, but see \citealt{cohenispi}) to a forthcoming study, 
we demonstrate below in
Sect.~\ref{bumphbsect} that an accurate characterization of the bump 
apparent magnitude, 
in combination with other features among luminous, evolved cluster members
such as the HB and TRGB, can yield 
distance- and reddening-independent cluster
metallicities with a useful precision. 

\subsubsection{Measuring the Bump Location}
\label{bumpfitsect}
To quantify the location of the RGB bump (RGBB) 
in our target GGCs and its uncertainty, 
we construct the LF of the RGB using 
only stars in sample A.    
In an attempt to maintain self-consistency in our analysis, the LF is built
with a binsize of 0.3 mag for all target clusters, although we
  found that the use of binsizes from 0.2-0.4 mag had a
  negligible effect on the resulting 
RGBB magnitudes compared to their uncertainties.  
To mitigate the
effects of binning, 10 histograms are constructed per cluster, but with the
bin starting points shifted fractionally each time by an increment 0.1 times 
the bin width, and the 10 histograms are then averaged
\citep[e.g.][]{bootstrap}.
 The resulting LF is then fit with an exponential
plus Gaussian \citep{natafrc,natafbump} as a function of apparent magnitude $m$ in each filter:

\begin{equation} 
  N(m) = A \exp[B(m-m_{RGBB})] +
  \frac{N_{RGBB}}{\sqrt{2\pi}\sigma_{RGBB}} \exp
  \left[-{\frac{(m-m_{RGBB})^2}{2\sigma^{2}_{RGBB}}}\right] 
\label{lfeq1} 
\end{equation}

\begin{figure}
\includegraphics[width=0.49\textwidth]{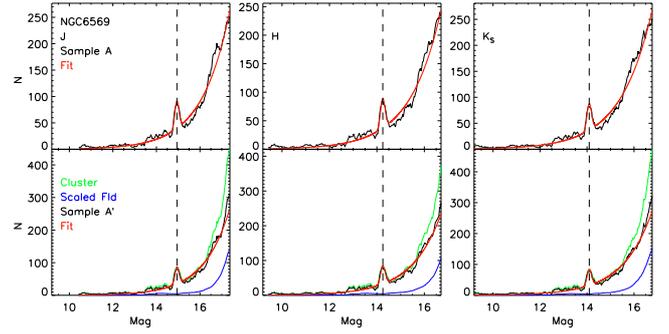}
\caption{An example of the observed 
LFs for NGC 6569 in all three $JHK_{S}$ filters (left to right).
In the upper panels, the LF constructed from the processed CMDs (Sample A)
is shown in black, and the exponential plus Gaussian fit obtained using
Eq.~\ref{lfeq1} is shown in red.
In the lower panels, we show the LFs constructed using Sample A$'$, in which
the LF is built using all stars in the cluster region (green), the LF of
the comparison field is constructed using an identical CMD region and scaled to
the spatial area of the cluster region (blue).  This scaled field LF is then 
subtracted to yield a field-subtracted cluster
LF (black), shown with the corresponding exponential plus Gaussian fit (red).
In all panels, the RGBB magnitude resulting from the fits 
is shown as a vertical dashed line.  RGB LFs as shown for this example case 
are included for all target clusters in the supplementary figures.}
\label{lfexam1}
\end{figure}

Here, $A$ is a scale factor, $B$ gives the 
exponential slope of the RGB, and $m_{RGBB}$ is the magnitude of the bump.  An
example of an observed LF and the resulting fit is shown in
Fig.~\ref{lfexam1}.

There are four cases where the HB intrudes on the RGB LF due to residual 
small-scale differential reddening which is unaccounted for by our maps.  In
these cases (NGC 6440, NGC 6441, NGC 6528 and NGC 6553), 
the HB causes a discernible second peak in
the RGB LF, 
so an exponential plus double Gaussian is fit \citep[e.g.][]{natafrc}: 

\begin{equation}
\begin{split} 
N(m) = A \exp[B(m-m_{RGBB})] + \frac{N_{RGBB}}{\sqrt{2\pi}\sigma_{RGBB}} \exp \left[-{\frac{(m-m_{RGBB})^2}{2\sigma^{2}_{RGBB}}}\right] \\ 
+ \frac{N_{HB}}{\sqrt{2\pi}\sigma_{HB}}\exp\left[(-{\frac{(m-m_{HB})^2}{2\sigma^{2}_{HB}}})\right] 
\end{split}
\label{lfeq2}
\end{equation}

\begin{figure}
\includegraphics[width=0.49\textwidth]{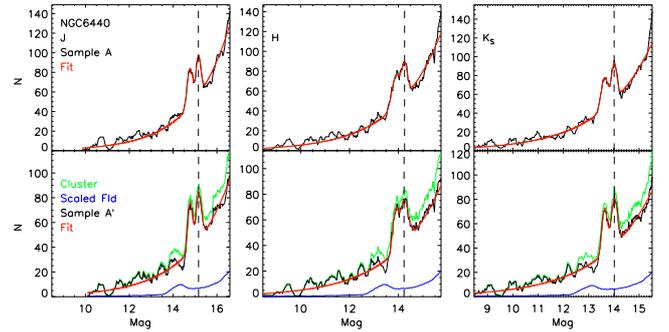}
\caption{As for Fig.~\ref{lfexam1}, but illustrating an exponential plus
  double Gaussian fit due to the intrusion of the HB on the RGB LF.}
\label{lfexam2}
\end{figure}

An example of an exponential plus double Gaussian fit is shown in
Fig.~\ref{lfexam2}.  

Because we employ the entire magnitude range of
the RGB for our exponential plus Gaussian fits rather than a restricted
magnitude range around the RGBB \citep[e.g.][]{natafbump,calamida6528}, 
the resulting RGBB magnitudes
are robust to both gaps in the LFs of the processed CMDs as well as stochastic
fluctuations at the bright end of the LFs due to the exponential nature of the
RGB LF.

\subsubsection{Quantifying Uncertainties}
\label{bumperrsect}
To calculate the total uncertainty on the
bump magnitudes 
resulting from the fit, 
we take our multi-binning approach as
well as the photometric errors into account using bootstrap resampling 
in each cluster.  For each of 1000 monte carlo iterations, all stars are offset
in the colour-magnitude diagram by a random amount drawn from a Gaussian
distribution that has a standard deviation equal to their photometric
error.  The entire fitting procedure is then repeated, including the multi-bin
generation of the LF and the exponential plus Gaussian fits, and the resulting
bump magnitudes are reported for each iteration.  To be conservative, the 
uncertainty which we report
for each parameter is the quadrature sum of the reported uncertainty from the
fit to the observed LF plus the standard deviation of the 1000 best-fitting 
values output
from the bootstrapping iterations.  
Furthermore, if the observed value of a
parameter is deviant from the median of the 1000 values output by the
bootstrapping procedure by more than this standard deviation, 
it is considered dubious, indicated by parentheses in Table \ref{obstab}.

\begin{figure}
\includegraphics[width=0.49\textwidth]{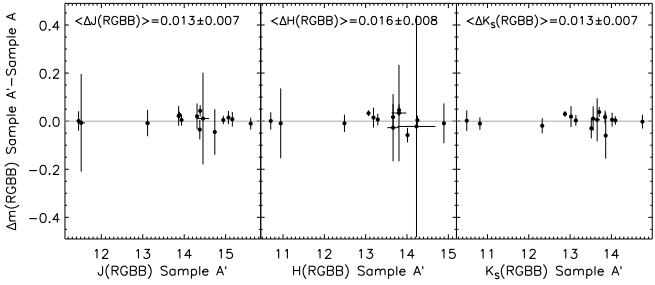}
\caption{The difference between the RGBB magnitudes measured using Sample A
  and Sample A$'$, shown as a function of $m(RGBB)$ from Sample A$'$ 
in all three
$JHK_{S}$ filters.  The grey horizontal line represents equality, and the mean
offset and its standard deviation are given at the top of each panel.}
\label{compbumpparams}
\end{figure}

To test whether the measurement of the RGBB magnitude is affected by
discontinuities or other artifacts of imperfect field star
  decontamination which may be present in the processed 
CMDs seen in Figs.~\ref{cmdsjk} and \ref{cmdsjh}, we have redetermined the
RGBB magnitudes using an alternate procedure.  Rather than constructing the LF
from the 
processed CMDs, we directly decontaminate the LF itself.  A multi-bin LF is
generated employing the stars in the CMD region occupied by Sample A, but
using all stars in the cluster region \textit{before} the decontamination
procedure was applied.  Next, another multi-bin LF is constructed from the
same CMD area, but using only stars physically located in the comparison
region (e.g.~outside the cluster tidal radii).
This comparison region LF is scaled to the
relative area of the cluster region and subtracted from the cluster region LF,
again performing 1000 monte carlo iterations where the comparison and cluster
stars are offset by Gaussian deviates of their
photometric errors.  A comparison between the RGBB magnitudes obtained from
this alternate procedure, which we refer to as Sample A$'$, versus those
obtained above from Sample A, is shown in Fig.~\ref{compbumpparams}.  
The mean offset in each filter between the RGBB
magnitude from Sample A and Sample A$'$ is given in each panel of
Fig.~\ref{compbumpparams} along with the standard deviation of the mean,
revealing a mean offset of $<$0.02 mag in all three filters.  
Furthermore, the uncertainties in the RGBB
determined using Sample A$'$ are not larger than those determined from Sample
A, and the ratio of the RGBB uncertainties measured from the two samples has a
median of 1 in all three filters.  Given the generally smoother LFs and more 
stable fits to sample A$'$ as compared to Sample A, we adopt the RGBB
magnitudes resulting from the fits to Sample A$'$.  However, both sets of RGB
LFs as presented in Figs.~\ref{lfexam1} and
\ref{lfexam2} are included for all target clusters 
in the supplementary figures, along with cluster
CMDs zoomed on the RGBs.  Finally, while a detailed
study of other RGBB parameters (i.e.~number counts, radial gradients, and
skewness) is better performed with high spatial resolution,
completeness-corrected photometry, the values of the LF
exponent $B$ that we obtain from Sample A$'$ are
$B$=(0.63,0.59)$\pm$(0.11,0.12) in $J$ and $K_{S}$ respectively, in reasonable
($<$1$\sigma$) agreement with values found in the $I$ band by
\citet{natafbump}.     

The magnitudes of the RGBB in each filter for all of our
target clusters are listed in Table \ref{obstab}, using parentheses to
indicate uncertain values.  The magnitudes
of photometric features given in Table \ref{obstab} are apparent magnitudes, 
measured using photometry which has
been corrected for reddening differentially across each cluster, but has not been 
corrected for total line-of-sight extinction or distance.

\begin{table*}
\caption{Observed RGBB, HB, and RGB Slope}
\begin{tabular}{lccccccccc}
Cluster & $\mathit{[Fe/H]}$ & $\mathit{[M/H]}$ & $J(RGBB)$ & $H(RGBB)$ & $K_{S}(RGBB)$ & $J(HB)$ & $H(HB)$ & $K_{S}(HB)$ & $Slope_{JK}$ \\
\hline
NGC6380$^{a}$ & -0.40$\pm$0.09 & -0.17$\pm$0.12 & 15.068$\pm$0.019 & 14.165$\pm$0.035 & 13.846$\pm$0.018 & 14.945$\pm$0.033 & 14.139$\pm$0.033 & 13.862$\pm$0.033 & -0.097$\pm$0.003 \\
NGC6401 & -1.01$\pm$0.14 & -0.76$\pm$0.16 & (14.16) & (13.31) & (13.14) &  &  &  & -0.076$\pm$0.002 \\
NGC6440$^{a}$ & -0.20$\pm$0.14 & 0.04$\pm$0.16 & 15.162$\pm$0.022 & 14.230$\pm$0.459 & 14.017$\pm$0.021 & 14.683$\pm$0.033 & 13.831$\pm$0.033 & 13.599$\pm$0.033 & -0.103$\pm$0.003 \\
NGC6441 & -0.44$\pm$0.07 & -0.29$\pm$0.10 & 15.605$\pm$0.018 & 14.893$\pm$0.022 & 14.751$\pm$0.021 & 15.125$\pm$0.031 & 14.483$\pm$0.031 & 14.346$\pm$0.031 & -0.104$\pm$0.002 \\
NGC6453 & -1.48$\pm$0.14 & -1.22$\pm$0.16 & 14.376$\pm$0.030 & 13.663$\pm$0.138 & 13.513$\pm$0.031 &  &  &  & -0.057$\pm$0.004 \\
NGC6522$^{a}$ & -1.45$\pm$0.08 & -1.20$\pm$0.11 & 13.866$\pm$0.027 & 13.189$\pm$0.028 & 13.023$\pm$0.029 &  &  &  & -0.083$\pm$0.002 \\
NGC6528$^{a}$ & 0.07$\pm$0.08 & 0.21$\pm$0.11 & 14.738$\pm$0.024 & 14.012$\pm$0.022 & 13.862$\pm$0.026 & 13.903$\pm$0.044 & 13.206$\pm$0.044 & 13.035$\pm$0.044 & -0.106$\pm$0.003 \\
NGC6544 & -1.47$\pm$0.07 & -1.21$\pm$0.11 & 11.451$\pm$0.028 & 10.702$\pm$0.025 & 10.499$\pm$0.030 &  &  &  & -0.072$\pm$0.005 \\
NGC6553 & -0.16$\pm$0.06 & 0.05$\pm$0.10 & 13.887$\pm$0.011 & 13.069$\pm$0.010 & 12.876$\pm$0.010 & 13.217$\pm$0.033 & 12.448$\pm$0.033 & 12.245$\pm$0.033 & -0.105$\pm$0.003 \\
NGC6558$^{a}$ & -1.37$\pm$0.14 & -1.10$\pm$0.16 & (13.78) & (13.16) & (13.02) &  &  &  & -0.082$\pm$0.003 \\
NGC6569$^{a}$ & -0.72$\pm$0.14 & -0.40$\pm$0.16 & 14.944$\pm$0.012 & 14.252$\pm$0.013 & 14.096$\pm$0.012 & 14.996$\pm$0.035 & 14.457$\pm$0.035 & 14.316$\pm$0.035 & -0.090$\pm$0.002 \\
NGC6624$^{a}$ & -0.42$\pm$0.07 & -0.15$\pm$0.11 & 14.309$\pm$0.040 & 13.662$\pm$0.040 & 13.558$\pm$0.038 & 13.918$\pm$0.033 & 13.411$\pm$0.033 & 13.315$\pm$0.033 & -0.104$\pm$0.003 \\
M28 & -1.46$\pm$0.09 & -1.20$\pm$0.12 & 13.116$\pm$0.051 & 12.486$\pm$0.029 & 12.326$\pm$0.025 &  &  &  & -0.081$\pm$0.002 \\
M69 & -0.59$\pm$0.07 & -0.37$\pm$0.10 & 14.385$\pm$0.016 & 13.808$\pm$0.016 & 13.704$\pm$0.014 & 14.051$\pm$0.032 & 13.560$\pm$0.032 & 13.485$\pm$0.032 & -0.098$\pm$0.003 \\
NGC6638 & -0.99$\pm$0.07 & -0.74$\pm$0.11 & 14.453$\pm$0.145 & 13.806$\pm$0.177 & 13.655$\pm$0.067 & 14.655$\pm$0.036 & 14.096$\pm$0.036 & 14.029$\pm$0.036 & -0.086$\pm$0.005 \\
NGC6642 & -1.19$\pm$0.14 & -0.94$\pm$0.16 & 13.936$\pm$0.016 & 13.292$\pm$0.017 & 13.140$\pm$0.016 & 14.455$\pm$0.061 & 14.125$\pm$0.061 & 14.047$\pm$0.061 & -0.075$\pm$0.004 \\
M22 & -1.70$\pm$0.08 & -1.47$\pm$0.11 & 11.514$\pm$0.086 & 10.943$\pm$0.021 & 10.818$\pm$0.017 &  &  &  & -0.067$\pm$0.002 \\
\hline
\multicolumn{8}{l}{$^{a}$ Cluster not used in calibrations due to uncertain
  metallicity, see Sect.~\ref{badfehsect}.}
\end{tabular}
\label{obstab}
\end{table*}

\subsection{Horizontal Branch Magnitude}
\label{hbsect}

Various methods have historically been applied to measure the 
magnitude of the HB and its uncertainty, 
including the median of a CMD-selected region \citep[e.g.][]{gs02,natafbump}, 
Gaussian fits to the LF peak \citep[e.g.][]{calamida6528}, 
and the location of the maximum of the cluster LF \citep[e.g][]{v04obs}.  
In the near-infrared, an obvious
complicating factor is the near-verticality
of the HB for less metal-rich clusters, so that an LF peak representative of
the HB location is not always detectable for GGCs with exclusively blue HBs
\citep[e.g.][]{cohenispi}.  Therefore, for compatibility with previous
studies, we restrict our HB analysis to clusters with relatively 
red HBs with a detectable peak in the LF of the HB, and
use the observed cluster LF peak to quantify the location of the HB in
$JHK_{S}$ magnitude.  In order to isolate the HB from the influence of
  the RGBB, we construct the LF
  using only stars in the processed CMDs in Sample B, which are those lying
  more than 3$\sigma$ blueward of the cluster fiducial sequences. 
The LF is built from this sample
using the
same binsizes, multi-binning, and bootstrap resampling as in the case of
the RGBB.  However, in lieu of a Gaussian fit to the LF, 
the reported HB magnitude is simply the magnitude corresponding to the
LF peak.  This is done both
for compatibility with previous near-infrared studies \citep[e.g.][]{v04obs}, 
and because models and data demonstrate that the HB LF may be non-Gaussian in 
near-IR magnitude \citep[e.g.][see their fig.~10]{47tucir}.  Therefore, as in
\citet{cohenispi}, the
reported uncertainties are the quadrature sum of the standard deviation of the
LF peak over the bootstrap iterations plus the 
effective resolution element of the LF.

We have performed our measurement of the HB LF and its peak neglecting known
RR Lyrae variables in our target clusters.  To check whether their inclusion
affects the measured HB magnitude, we have reperformed our fits (including the
bootstrapping iterations) with all known variables included.  We found that in
all cases the resultant HB magnitude is unaffected beyond the reported
uncertainties, consistent with simulations by \citet{milonehb} 
demonstrating that
even in optical bandpasses the influence of RRL photometric variability 
on single-epoch photometry negligibly affected the HB morphological 
parameters which they measured.

\begin{figure}
\includegraphics[width=0.49\textwidth]{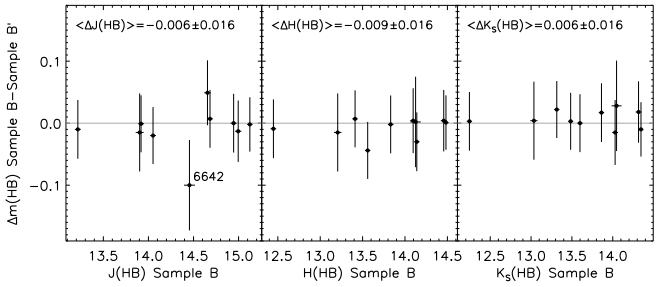}
\caption{The difference between the HB magnitudes measured using Sample B
  and Sample B$'$, shown as a function of $m(HB)$ from Sample B 
in all three
$JHK_{S}$ filters.  Symbols are as in Fig.~\ref{compbumpparams}.}
\label{comphbparams}
\end{figure}

In order to assess the influence of the decontamination procedure, including
potential imperfect subtraction of blue Galactic disk stars on the
measured HB magnitudes, we have performed a comparison analogous to 
Fig.~\ref{compbumpparams}.  Specifically, the HB LF was generated using all
stars in the CMD region occupied by Sample B \textit{before} statistical
decontamination, and a field HB LF was generated from this same CMD area
using stars spatially located in the comparison region.  The comparison region
LF was scaled to the area of the cluster region and subtracted before
measuring the peak of the resultant LF over 1000 bootstrapping iterations in
which photometric errors were applied to both the cluster and comparison
region stars.  A comparison between the HB magnitudes measured from this
sample, denoted as Sample B$'$, and the HB magnitudes measured from Sample B
(using the statistically decontaminated CMD directly) is shown in
Fig.~\ref{comphbparams}.  This comparison illustrates that the HB magnitudes
obtained using the two methods agree to within their uncertainties, with the 
only slight ($<$1.4$\sigma$) exception of NGC 6642 in the $J$ band, which in
any case is excluded from the calibration of our photometric metallicity
relations (see Sect.~\ref{photcalsect}).

\subsection{Comparison With Literature Values}
\label{comparsect}
We compare our observed values listed in Tables \ref{obstab} and Table \ref{tipstartab} 
with those from the
literature, using the most recent sources as follows: Where available, we use 
values from the 
systematic near-IR photometric studies of our target clusters by
\citet{v04obs,v04abs} and \citet{v10}.  
Otherwise we take from \citet{chun} the $K_{S}$ values and RGB
slope for NGC 6642, the $K_{S}$
magnitude of the RGB bump for NGC 6401 and all available near-IR parameters
for M 28.  Additionally, the magnitude of the RGB tip in M 22
from 2MASS is taken from \citet{monaco}.
In Table \ref{comptab},
we list the mean offset between our values and
these literature values and its standard deviation, 
as well as the total number of clusters available for
comparison.  Bearing in mind that both photometric calibration
uncertainties as well as observational measurement uncertainties 
contribute to this difference, the values are generally in
good agreement.  Our values for the TRGB magnitude lie $\sim$0.1 faintward of
those reported by \citet{v04abs}, consistent with the suggestions of both
\citet{dalcanton} and \citet{marektrgb} that the near-IR TRGB magnitude from
the \citet{v04abs} calibration is 0.1-0.2 mag too bright.
In the
present case, the discrepancy could be partially due to the exclusion of 
(then-unknown) AGB variables, although it is well within the margin suggested
by measurement error alone: The median
published uncertainty of literature TRGB measurements is 0.22 mag in $K_{S}$ 
and 0.23 mag in $J$ and $H$ (noting that these reported values neglect the 
additional contribution from uncertainties in the photometric calibration to 
the 2MASS system), and we revisit empirical constraints on the precision of
the near-IR TRGB magnitude in Sect.~\ref{bumptipsect}. 

\begin{table}
\centering
\caption{Comparison to Literature Values}
\begin{tabular}{lcc}
 Parameter & $\langle$This Study-Literature$\rangle$ & N(clus) \\ 
\hline
$J(RGBB)$ &  -0.045$\pm$0.020 & 10 \\
$H(RGBB)$ &   0.014$\pm$0.027 & 7 \\ 
$K_{S}(RGBB)$ &  -0.030$\pm$0.020 & 11 \\ 
$J(HB)$ & -0.052$\pm$0.020 & 9 \\
$H(HB)$ & 0.004$\pm$0.023 & 7 \\ 
$K_{S}(HB)$ & -0.025$\pm$0.020 & 7 \\ 
$J(TRGB)$ &   0.102$\pm$0.026 & 14 \\ 
$H(TRGB)$ &   0.113$\pm$0.028 & 11 \\ 
$K_{S}(TRGB)$ &   0.117$\pm$0.043 & 16 \\ 
Slope$_{JK}$ & -0.0003$\pm$0.0011 & 11 \\ 
\hline
\end{tabular}
\label{comptab}
\end{table}

\subsection{Some Special Cases}

There are a few specific cases of clusters for which a single HB/RGBB value
may not be appropriate that deserve some mention.  For the double
HBs of NGC 6440 and NGC 6569 reported by \citet{maurohb}, the values which we obtain are in good agreement, both
intermediate between the two HB peaks which they report in each cluster: For 
NGC 6440, we find $K_{S}(HB)$=13.599$\pm$0.033, in comparison with their
values of 13.55 and 13.67 for the two HBs, and for NGC 6569, we obtain
$K_{S}(HB)$=14.316$\pm$0.035 in comparison to 14.26 and 14.35 for the two HBs.
Since we employ the same data as in that study, 
we cannot constrain the nature of
the HBs beyond the results which they report, and a more detailed study of the
HB morphology in these clusters using deep, high resolution imaging is
underway (F.~Mauro et al.~in prep.).  
Our $K_{S}(HB)$ values also agree well with those employed by \citet{mauro14} to
devise reduced CaII equivalent width-$\mathit{[Fe/H]}$ relations: 
All clusters in
common have $K_{S}(HB)$ values which agree to within their uncertainties, with 
the exception of NGC 6638, for which \citet{mauro14} report a significantly 
brighter value (13.70$\pm$0.05 versus 14.029$\pm$0.036).  
This merely reflects a
difference in methodology, since \citet{mauro14} used the reddest 
part of the HB, as given by theoretical models in combination with distances of
\citet{v10}, to
calculate their $K_{S}(HB)$ values, whereas we report the location of the 
observed LF peak.  

We also compare our $K_{S}(HB)$ and $K_{S}(RGBB)$ values to those
reported for NGC 6528 by 
\citet{calamida6528} using a sample of
proper motion selected cluster members.  
Their value of $K_{S}$(bump)=13.85$\pm$0.05 compares well with our
measurement of $K_{S}$(bump)=13.862$\pm$0.026, and they suggest a double-peaked HB with peaks at 
$K_{S}$=12.97$\pm$0.02 and 13.16$\pm$0.02.
As we employ some of the
same data which they used, we cannot comment further on this feature, but our
intermediate value of $K_{S}(HB)$=13.035$\pm$0.044 
supports both the location and
atypically large width in magnitude which they report 
for the HB of this cluster.  As they cite possible residual field
contamination of their proper-motion-selected sample as one possible cause 
of the bimodality, a detailed study of this feature may benefit from 
high resolution near-infrared imaging of a thoroughly cleaned sample
of cluster members (R.~E.~Cohen et al., in prep.).

\section{Distance- and Reddening-Independent Calibrations}
\label{photcalsect}
Several of the photometric parameters which we have reported can be used to
construct indices from
relative measurements made on a cluster CMD.  By choosing a set of calibrating
clusters with well-measured metallicities, we can build relations between
photometric indices versus metallicity which may be applied as distance- and
reddening-independent metallicitity indicators.  The relative photometric
indices which we explore include the slope of the RGB in the
$K_{S},(J-K_{S})$ plane ($slope_{JK}$), as well as the magnitude difference 
between the HB and RGBB ($\Delta$$m^{HB}_{RGBB}$) and the magnitude
difference between the RGB bump and the tip of the RGB ($\Delta$$m^{RGBB}_{TRGB}$) 
in each of the three $JHK_{S}$ bandpasses.  
While we defer a discussion of calibrations versus
\textit{absolute} magnitude, and hence the GGC distance scale, to a forthcoming
publication, the distance- and reddening-independent 
relations which we derive can serve as quantitative tests
of evolutionary models as well as photometric metallicity indicators for
old stellar populations. 

\subsection{Input Metallicities:$\mathit{[Fe/H]}$}
\label{badfehsect}
\begin{figure}
\includegraphics[width=0.49\textwidth]{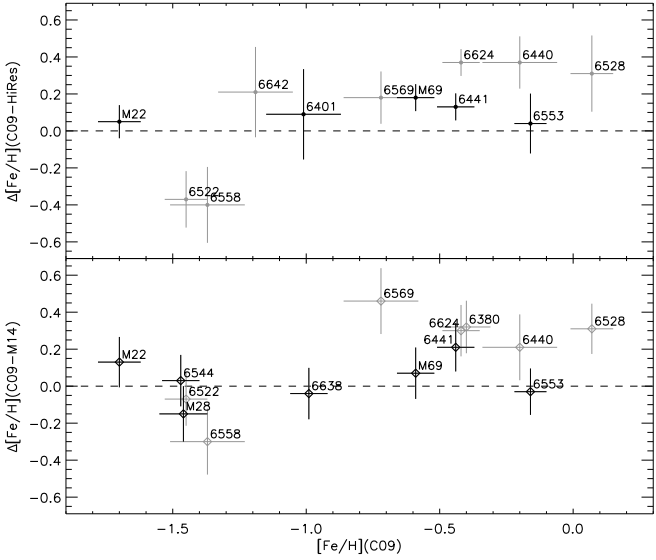}
\caption{Comparison between $\mathit{[Fe/H]}$ values reported by \citet{c09},
  versus those from independent spectroscopic studies (top panel) as well
as the CaII triplet calibration of \citet{mauro14} (bottom panel).  Clusters
included as calibrators are plotted in black, and those excluded due to
controversial $\mathit{[Fe/H]}$ values are plotted in grey.  The dotted horizontal line
in each panel represents equality, and clusters are labelled by NGC or Messier number.}
\label{compfeh}
\end{figure}

We wish to use GGCs with the most reliable spectroscopic abundances to
calibrate relations between photometric features and cluster metallicity.
While recent
large-scale spectroscopic campaigns have vastly increased the number of GGCs 
with high-quality self-consistent spectroscopic measurements of both 
$\mathit{[Fe/H]}$ and $[\alpha/Fe]$ \citep[][]{c09,c10,brunolores},  
the issue of spectroscopic metallicities
remains complicated with regard to the GGCs located towards the
Galactic bulge.  In some cases, the values of $\mathit{[Fe/H]}$ listed by \citet{c09}
are significantly at odds with those from other recent, independent
spectroscopic investigations.  Therefore, we summarize in Table \ref{fehtab}
various spectroscopic $\mathit{[Fe/H]}$ values for our target clusters from several
sources in addition to \citet{c09}. These include the 
near-infrared CaII triplet studies by 
\citet{mauro14},
and an additional set comprised of any independent spectroscopic metallicity
measurements in the literature, which we denote as ``HiRes''.  These values are 
further compared in  
Fig.~\ref{compfeh}, where we plot $\mathit{[Fe/H]}$ from \citet{c09} versus the HiRes
values in the top panel.  Significant ($>$0.3 dex) discrepancies are evident,
as noted by
\citet{mauro14}, who devised a set of ``corrected'' \citet{c09}
values (which they denote ``C09c'') 
for clusters where \citet{c09} values showed 
significant discrepancies from other studies.  
In the bottom panel of Fig.~\ref{compfeh},
we compare \citet{c09} $\mathit{[Fe/H]}$ with the values given by the CaII triplet
calibrations of \citet{mauro14}, employing their best-fitting
relations of CaII equivalent width versus C09c $\mathit{[Fe/H]}$ values, 
which are cubic in the
case of the \citet{saviane} equivalent widths (column IIIa of their table 3) 
and quadratic in the case of the
Rutledge, Hesser \& Stetson (1997) 
equivalent widths (column IIa of their table 6).  
The uncertainties on the \citet{mauro14} $\mathit{[Fe/H]}$ values employed in Table
\ref{fehtab} and Fig.~\ref{compfeh} are the unbiased rms which they report
from the applicable calibration (evaluated 
considering only the calibrating clusters), and for the two clusters with
equivalent width measurements from both \citet{saviane} and \citet{r97} 
(NGC 6528 and NGC 6553), we use the $\mathit{[Fe/H]}$ values
resulting from the calibration employing the more recent 
\citet{saviane} equivalent widths.

Given the evident discrepancies in $\mathit{[Fe/H]}$ for some clusters, we adopt the
following strategy: Clusters with controversial $\mathit{[Fe/H]}$ values, 
plotted in
grey in Fig.~\ref{compfeh}, are excluded from our calibrations of photometric
indices versus metallicity, and we later
use our results to comment on the metallicities of these clusters.  The
remainder of our VVV targets, plotted in black in Fig.~\ref{compfeh}, are
used to calibrate photometric metallicity indicators, together with recent
literature results (see below).  
Meanwhile, a few of the cases listed in Table \ref{fehtab}
deserve further comment regarding their HiRes $\mathit{[Fe/H]}$ values as they have
been subjected to multiple recent spectroscopic investigations:

\noindent\textbf{NGC 6522:} Spectroscopic 
analyses were recently presented by both Ness, Asplund \& Casey (2014) and \citet{barbuy14}, 
targeting eight and four giants respectively.  
Despite having several stars in common, the two studies report
mean $\mathit{[Fe/H]}$ values which differ by 0.2 dex, albeit with an uncertainty of 0.15
dex in both cases.  In addition, \citet{ness} find that the cluster is 
significantly
$\alpha$-enhanced while \citet{barbuy14} claim only low to moderate
enhancements of Si, Ca and Ti.  We choose to adopt the abundances of
\citet{ness} due to their larger sample size, but we
recalculate the mean $\mathit{[Fe/H]}$ excluding star B-108 as \citet{barbuy14} found 
that it is blended, bringing the two studies into agreement at the 
1$\sigma$ level.  
While we cannot exclude the possibility that any of the
three stars in the \citet{ness} sample not studied by \citet{barbuy14} is
likewise affected by blending, there is no obvious indication among the reported
radial velocities or abundances that this is the case.  

\noindent\textbf{NGC 6528:} As discussed by \citet{mauro14} and \citet{brunolores}, several recent
spectroscopic studies have found $\mathit{[Fe/H]}$ values lower than the super-solar
value of $\mathit{[Fe/H]}$=+0.07$\pm$0.08 from \citet{carretta01} used in the
compilation of \citet{c09}.  
\citet{origlia05} report $\mathit{[Fe/H]}$=-0.17$\pm$0.01 from high resolution near-IR
spectra of four RGB stars, while \citet{zoccali04} and \citet{sobeck06} report
$\mathit{[Fe/H]}$=-0.1$\pm$0.2 and -0.24$\pm$0.19 dex respectively from high resolution
optical spectra of three stars (one HB star and two RGB stars).  Both of these
values are in good agreement with the low-resolution optical spectra of 17 stars by 
\citet{dias14}, who report $\mathit{[Fe/H]}$=-0.13$\pm$0.05, and we adopt the estimate
of \citet{sobeck06} for the HiRes set of $\mathit{[Fe/H]}$ values and
comment further on photometric constraints in Sect.~\ref{discusssect}.

\noindent\textbf{M 22 (NGC 6656):} Several spectroscopic investigations have
claimed a split/multimodality in $\mathit{[Fe/H]}$
\citep{marino11,marino12,alvesbrito,marino13}.  However,
\citet{mucciarellim22} found that when FeI lines, which are more vulnerable to
non-local thermodynamic equilibrium effects, are excluded, FeII lines show no
significant spread in iron abundance.  
For our purposes, this turns out to be somewhat of a moot point, since the
value which \citet{mucciarellim22} calculate from FeII lines,
$[FeII/H]$=-1.75$\pm$0.04, is in good agreement with the \citet{c09}
value of $\mathit{[Fe/H]}$=-1.70$\pm$0.08, so we include this cluster in our set of
calibrators.\footnote{Incidentally, these $\mathit{[Fe/H]}$ values are not, \textit{on
    average}, inconsistent with other high-resolution studies.  
    Despite reporting a bimodality in $\mathit{[Fe/H]}$, the spectroscopic study of 35 RGB
    stars by \citet{marino11} gives a  
    \textit{mean} value of $\mathit{[Fe/H]}$=-1.77$\pm$0.03 dex.}      

\subsection{The Global Metallicity $\mathit{[M/H]}$}

Since models and observations both suggest that the upper RGB is sensitive to
variations in $[\alpha/Fe]$ as well as $\mathit{[Fe/H]}$
in the near-IR \citep{cohenispi}, we build relations
in terms of both $\mathit{[Fe/H]}$ and the global metallicity $\mathit{[M/H]}$, defined by 
\citet{salaris} as $[M/H]=[Fe/H]+Log (0.638f_{\alpha}+0.362)$, where
$f_{\alpha}=10^{[\alpha/Fe]}$ .
For clusters without spectroscopic measurements of $[\alpha/Fe]$, we assume
the linear $[\alpha/Fe]$-$\mathit{[Fe/H]}$ relation of \citet{natafbump}, and
conservatively assume $\sigma$$[\alpha/Fe]$=0.1 dex.  This yields
0.3$<$$[\alpha/Fe]$$<$0.4 for all target clusters without literature values of
$[\alpha/Fe]$, in accord with spectroscopic measurements of $\alpha$-enhancement
found in GGCs towards the Galactic bulge
\citep{origlia02,origlia05,barbuy07,OVR08,VOR11,valentiter1}.  In the case of NGC 6528, \citet{zoccali04} find $[\alpha/Fe]$$\sim$0.1, whereas \citet{origlia05} report $[\alpha/Fe]$$\sim$0.33, so we assume $[\alpha/Fe]$=0.2$\pm$0.1.
In all
cases, we 
calculate the uncertainty in the resulting $\mathit{[M/H]}$ following eq.~7
of \citet{natafbump}\footnote{Where not given explicitly, 
we calculate $\alpha$ as the mean of Ti, Si, Mg and Ca (e.g.~\citealt{VOR11}) 
weighted by the inverse squares of their uncertainties.}, 
and the values of $[\alpha/Fe]$ and their sources as well as the
resulting $\mathit{[M/H]}$ for our VVV target clusters are listed in Table \ref{fehtab}.

\begin{table*}
\centering
\caption{$\mathit{[Fe/H]}$ and $\mathit{[M/H]}$ Values for Target Clusters}
\begin{tabular}{lccccccc}
\hline
Cluster & $\mathit{[Fe/H]}$\citep{c09} & $[\alpha/Fe]$ & Reference & $\mathit{[M/H]}$
  & $\mathit{[Fe/H]}$\citep{mauro14} & $\mathit{[Fe/H]}$(HiRes) &
  Reference \\
\hline
NGC6380$^{a}$ & -0.40$\pm$0.09 & & & -0.17$\pm$0.12 & -0.72$\pm$0.11 & & \\
NGC6401 & -1.01$\pm$0.14 & & & -0.76$\pm$0.16 & & -1.10$\pm$0.20 & 1 \\
NGC6440$^{a}$ & -0.20$\pm$0.14 & 0.34 & 2 & 0.04$\pm$0.16 & -0.41$\pm$0.11 &
-0.57$\pm$0.02 & 2 \\
NGC6441 & -0.44$\pm$0.07 & 0.21 & 2 & -0.29$\pm$0.10 & -0.65$\pm$0.11 &
-0.57$\pm$0.02 & 2 \\
NGC6453 & -1.48$\pm$0.14 & & & -1.22$\pm$0.16 & & & \\
NGC6522$^{a}$ & -1.45$\pm$0.08 & 0.35 & 3 & -1.20$\pm$0.11 & -1.38$\pm$0.12 &
-1.08$\pm$0.13 & 3 \\
NGC6528$^{a}$ & 0.07$\pm$0.08 & 0.20 & 4,5 & 0.21$\pm$0.11 & -0.24$\pm$0.11 &
-0.24$\pm$0.19 & 6 \\
NGC6544 & -1.47$\pm$0.07 & & & -1.21$\pm$0.11 & -1.50$\pm$0.12 & & \\
NGC6553 & -0.16$\pm$0.06 & 0.30 & & 0.05$\pm$0.10 & -0.13$\pm$0.11 &
-0.20$\pm$0.15 & 7 \\
NGC6558$^{a}$ & -1.37$\pm$0.14 & 0.37 & 8 & -1.10$\pm$0.16 & -1.07$\pm$0.11 &
-0.97$\pm$0.15 & 8 \\
NGC6569$^{a}$ & -0.72$\pm$0.14 & 0.43 & 9 & -0.40$\pm$0.16 & -1.18$\pm$0.11 &
-0.90$\pm$0.02 & 9 \\
NGC6624$^{a}$ & -0.42$\pm$0.07 & 0.37 & 9 & -0.15$\pm$0.11 & -0.72$\pm$0.12 &
-0.79$\pm$0.02 & 9 \\
M28 & -1.46$\pm$0.09 & & & -1.20$\pm$0.12 & -1.31$\pm$0.12 & & \\
M69 & -0.59$\pm$0.07 & 0.31 & 10 & -0.37$\pm$0.10 & -0.66$\pm$0.12 &
-0.77$\pm$0.02 & 10 \\
NGC6638 & -0.99$\pm$0.07 & & & -0.74$\pm$0.11 & -0.95$\pm$0.12 & & \\
NGC6642$^{a}$ & -1.19$\pm$0.14 & & & -0.94$\pm$0.16 & & -1.40$\pm$0.20 & 1 \\
M22 & -1.70$\pm$0.08 & 0.32 & 11 & -1.47$\pm$0.11 & -1.83$\pm$0.11 &
-1.75$\pm$0.04 & 12 \\
\hline
\multicolumn{8}{l}{$^{a}$ Cluster excluded from calibrations in Sections
  \ref{slopesect}, \ref{bumphbsect} and \ref{bumptipsect} due to uncertain metallicity.  
  References: (1)\citealt{minniti95}}\\
\multicolumn{8}{l}{
  (2)\citealt{OVR08} 
 (3)\citealt{ness} 
(4)\citealt{zoccali04}
(5)\citealt{origlia05}
(6)\citealt{sobeck06}}\\
\multicolumn{8}{l}{
(7)\citealt{alvesbrito06}
 (8)\citealt{barbuy07}
(9)\citealt{VOR11}
(10)\citealt{lee07}
(11)\citealt{marino11}}\\
\multicolumn{8}{l}{
(12)\citealt{mucciarellim22}}
\end{tabular}
\label{fehtab}
\end{table*}

\subsection{Extending the Calibration Baseline}
\label{baselinesect}
Although our target clusters span a reasonably broad range in metallicity, 
they suffer from the limitation that there are no GGCs included which are more
metal-poor than M 22 ($[Fe/H]\lesssim$-1.7).  
In order to maximize the applicable metallicity range of our calibrations as
well as increase the sample size, we supplement the values which
we measure with those available in the literature which have high-quality
spectroscopic abundances \citep{c09,c10}.  
For this purpose, we denote the target clusters
described thus far as the ``VVV'' sample (including NGC 6544, as described in \citealt{cohen6544}), and supplement them with
near-IR measurements of 12 optically well-studied GGCs from \citet{cohenispi}
(the ``ISPI'' sample), as well
as the database of near-IR GGC photometry from \citet{v10} 
and references therein,
designated the ``V10'' sample\footnote{For values listed without uncertainties
  by \citet{v10}, we conservatively assume an uncertainty
of 0.005 on the RGB slope and 0.1 mag on the HB and RGB bump magnitudes based on
the errors reported by \citet{v04obs,v04abs}.}.  For convenience, we have
compiled measured photometric features from these supplementary sources in 
Table \ref{litparamtab}.

\begin{landscape}
\begin{table}
\caption{Literature Values of RGBB, HB, and RGB Slope}
\begin{tabular}{lcccccccccc}
Cluster & $\mathit{[Fe/H]}$ & $\mathit{[M/H]}$ & $J(RGBB)$ & $H(RGBB)$ &
$K_{S}(RGBB)$ & $J(HB)$ & $H(HB)$ & $K_{S}(HB)$ & $Slope(JK)$ & Reference \\
\hline
NGC104 & -0.76$\pm$0.02 & -0.47$\pm$0.08 & 12.728$\pm$0.010 & 12.15$\pm$0.05$^a$ & 12.072$\pm$0.009 & 12.492$\pm$0.013 & 12.02$\pm$0.04$^b$ & 11.979$\pm$0.013 & -0.098$\pm$0.001 & ISPI \\
NGC288 & -1.32$\pm$0.02 & -1.03$\pm$0.08 & 13.736$\pm$0.035 & 13.25$\pm$0.05$^a$ & 13.150$\pm$0.034 &  &  &  & -0.084$\pm$0.003 & ISPI \\
NGC362 & -1.30$\pm$0.04 & -1.01$\pm$0.09 & 13.754$\pm$0.026 & 13.25$\pm$0.05$^a$ & 13.136$\pm$0.029 & 14.155$\pm$0.011 & 13.775$\pm$0.032$^b$ & 13.680$\pm$0.011 & -0.078$\pm$0.002 & ISPI \\
NGC1261 & -1.27$\pm$0.08 & -0.98$\pm$0.11 & 14.997$\pm$0.009 &  & 14.420$\pm$0.009 & 15.418$\pm$0.022 &  & 14.928$\pm$0.026 & -0.079$\pm$0.003 & ISPI \\
NGC1851 & -1.18$\pm$0.08 & -0.89$\pm$0.11 & 14.424$\pm$0.025 &  & 13.808$\pm$0.028 & 14.775$\pm$0.012 &  & 14.340$\pm$0.011 & -0.084$\pm$0.002 & ISPI \\
NGC2808 & -1.18$\pm$0.04 & -0.89$\pm$0.09 & 14.253$\pm$0.013 &  & 13.530$\pm$0.012 &  &  &  & -0.085$\pm$0.001 & ISPI \\
NGC4833 & -1.89$\pm$0.05 & -1.60$\pm$0.10 & 12.917$\pm$0.014 &  & 12.133$\pm$0.028 &  &  &  & -0.055$\pm$0.002 & ISPI \\
NGC5927 & -0.29$\pm$0.07 & -0.15$\pm$0.10 & 14.585$\pm$0.027 &  & 13.760$\pm$0.038 & 14.036$\pm$0.013 &  & 13.254$\pm$0.011 & -0.101$\pm$0.004 & ISPI \\
NGC6304 & -0.37$\pm$0.07 & -0.23$\pm$0.10 & 14.211$\pm$0.017 & 13.33$\pm$0.10$^a$ & 13.295$\pm$0.029 & 13.577$\pm$0.022 & 12.85$\pm$0.05$^a$ & 12.709$\pm$0.024 & -0.097$\pm$0.004 & ISPI \\
NGC6496 & -0.46$\pm$0.07 & -0.32$\pm$0.10 & 14.783$\pm$0.024 &  & 14.015$\pm$0.035 & 14.288$\pm$0.036 &  & 13.666$\pm$0.030 & -0.100$\pm$0.002 & ISPI \\
NGC6584 & -1.50$\pm$0.09 & -1.21$\pm$0.12 & 14.624$\pm$0.011 &  & 13.999$\pm$0.010 &  &  &  & -0.070$\pm$0.001 & ISPI \\
NGC7099 & -2.33$\pm$0.02 & -2.04$\pm$0.08 &  &  &  &  &  &  & -0.049$\pm$0.002 & ISPI \\
NGC4590 & -2.27$\pm$0.04 & -2.02$\pm$0.06 & 13.35$\pm$0.05 &  & 12.80$\pm$0.05 &  &  &  & -0.048$\pm$0.003 & V10 \\
NGC5272 & -1.50$\pm$0.05 & -1.26$\pm$0.07 & 13.70$\pm$0.05 &  & 13.10$\pm$0.05 &  &  &  & -0.071$\pm$0.003 & V10 \\
NGC5904 & -1.33$\pm$0.02 & -1.05$\pm$0.05 & 13.25$\pm$0.05 &  & 12.65$\pm$0.05 &  &  &  & -0.082$\pm$0.004 & V10 \\
NGC6171 & -1.03$\pm$0.02 & -0.66$\pm$0.05 & 13.25$\pm$0.05 &  & 12.50$\pm$0.05 &  &  &  & -0.075$\pm$0.005 & V10 \\
NGC6205 & -1.58$\pm$0.04 & -1.36$\pm$0.06 & 13.05$\pm$0.05 &  & 12.40$\pm$0.05 &  &  &  & -0.065$\pm$0.002 & V10 \\
NGC6273 & -1.76$\pm$0.07 & -1.49$\pm$0.09 & 13.65$\pm$0.10 & 13.05$\pm$0.10 & 12.85$\pm$0.10 &  &  &  & -0.063$\pm$0.005 & V10 \\
NGC6293 & -2.01$\pm$0.14 & -1.70$\pm$0.15 &  &  &  &  &  &  & -0.048$\pm$0.005 & V10 \\
NGC6316 & -0.36$\pm$0.14 & -0.13$\pm$0.15 & 15.20$\pm$0.10 & 14.65$\pm$0.10 &  & 14.93$\pm$0.10 & 14.25$\pm$0.10 &  &  & V10 \\
NGC6341 & -2.35$\pm$0.05 & -2.01$\pm$0.07 & 12.85$\pm$0.05 &  & 12.35$\pm$0.05 &  &  &  & -0.046$\pm$0.003 & V10 \\
NGC6342 & -0.49$\pm$0.14 & -0.21$\pm$0.15 & 14.65$\pm$0.10 & 13.85$\pm$0.10 & 13.75$\pm$0.10 & 14.25$\pm$0.05 & 13.60$\pm$0.05 & 13.40$\pm$0.05 & -0.102$\pm$0.003 & V10 \\
NGC6355 & -1.33$\pm$0.14 & -1.07$\pm$0.15 &  &  &  &  &  &  & -0.068$\pm$0.005 & V10 \\
NGC6388 & -0.45$\pm$0.04 & -0.30$\pm$0.06 & 15.18$\pm$0.10 & 14.47$\pm$0.10 & 14.33$\pm$0.10 & 14.90$\pm$0.10 & 14.27$\pm$0.10 & 14.17$\pm$0.10 &  & V10 \\
NGC6539 & -0.53$\pm$0.14 & -0.21$\pm$0.15 & 14.90$\pm$0.10 & 14.05$\pm$0.10 & 13.83$\pm$0.10 & 14.65$\pm$0.10 & 13.85$\pm$0.10 & 13.65$\pm$0.10 &  & V10 \\
NGC6752 & -1.55$\pm$0.01 & -1.23$\pm$0.05 & 11.90$\pm$0.05 & 11.35$\pm$0.05 & 11.25$\pm$0.05 &  &  &  & -0.048$\pm$0.003 & V10 \\
NGC6809 & -1.93$\pm$0.02 & -1.62$\pm$0.05 & 12.35$\pm$0.05 &  & 11.75$\pm$0.05 &  &  &  & -0.049$\pm$0.003 & V10 \\
NGC7078 & -2.33$\pm$0.02 & -2.00$\pm$0.05 & 13.55$\pm$0.05 & 13.05$\pm$0.05 & 12.95$\pm$0.05 &  &  &  & -0.044$\pm$0.003 & V10 \\
\hline
\multicolumn{11}{l}{$^{a}$ Where available, $H(RGBB)$ and $H(HB)$ for ISPI
  clusters taken from \citet{v10} and references therein.} \\
\multicolumn{11}{l}{$^{b}$ $H(HB)$ for NGC104 from fig.~10 of
  \citet{47tucir} and $H(HB)$ for NGC362 calculated from 2MASS photometry as described in \citet{cohenispi}.}
\end{tabular}
\label{litparamtab}
\end{table}
\end{landscape}

We now describe the construction of relations between cluster metallicity versus
several relative photometric indices which can be measured from cluster CMDs.  
The indices studied typically span a colour range of
$\Delta(J-K_{S})\lesssim$0.5, so in addition to being independent of distance 
and reddening, they are insensitive  
to photometric zeropoint uncertainties and the assumed reddening law.   

\subsection{Red Giant Branch Slope ($slope_{JK}$)}
\subsubsection{Observed Slope Measurements}
\label{slopesect}

A linear relation has traditionally been used to describe cluster metallicity
(in terms of $\mathit{[Fe/H]}$ and/or $\mathit{[M/H]}$) versus the slope of the upper RGB (in
terms of colour as function of magnitude), 
calculated over a magnitude range on the upper RGB 
where the effects of metallicity variations are most prominent
\citep[e.g.][]{v04obs}. 
 The definition of this magnitude range is based on the
observation of \citet{kuchinski} that using stars in the range 0.6 to 5.1 mag 
brighter than the zero age horizontal branch (ZAHB) serves to avoid the
influence of HB stars at the faint end and bright AGB variables close to the
RGB tip.  However, defining a ZAHB magnitude for metal-poor clusters in the
near-IR is difficult since their horizontal branches are in fact
almost vertical, so for consistency we follow the methodology of \citet[][and
references therein]{v04obs} and use the magnitude range 
0.5$<$$(K-K_{TRGB})$$<$5.0.   
The RGB slope is measured by fitting a line to all stars
which lie in this magnitude range and have colours within 3$\sigma$ of the
fiducial sequence (e.g. included in Sample A).

In Fig.~\ref{slopefig}, we show measured RGB slope values versus both
$\mathit{[Fe/H]}$ and $\mathit{[M/H]}$ (on the \citet{c09} scale).  
Target clusters from the VVV sample are shown as filled black circles, while
those from the literature are shown using blue squares (V04 sample) or red
circles (ISPI sample).  VVV clusters which were not used as calibrators due to
their uncertain $\mathit{[Fe/H]}$ values are shown in grey and labelled by NGC number, 
and for each of these clusters, a
vertical dotted line is shown
connecting their \citet{c09} $\mathit{[Fe/H]}$ values to those from the \citet{mauro14}
and HiRes scales as reported in Table \ref{fehtab} (for clarity, error bars on
these alternate values are not shown in Fig.~\ref{slopefig}).  As in the case
of the HB and RGBB magnitudes, the reported slope errors are the quadrature
sum of the formal uncertainty on the slope from an unweighted least squares fit (see below) 
plus the standard deviation of the slopes obtained over 1000 bootstrapping
iterations in which the stars are offset by Gaussian deviates of their
photometric errors.  Given the difficulties in separating AGB and RGB cluster 
members
discussed in Sect.~\ref{tipsect}, we have chosen to exclude all known variables
from our slope measurements, and the use of a colour cut combined with the
exclusion of stars within 0.5 mag of the TRGB serves to effectively 
remove most known variables from the CMD region used to measure the slope.  
However, in the minority of cases where a
small number of variables fall in the CMD region used for slope measurement
(1, 2 and 5 stars each in M28, M22 and NGC6441 respectively),
we have reperformed the slope measurement including these variables and
verified that the slopes are unaffected beyond their uncertainties.  

\subsubsection{Uncertainties in the Slope Measurement}
\label{slopeerrsect}
\begin{figure*}
\includegraphics[width=0.97\textwidth]{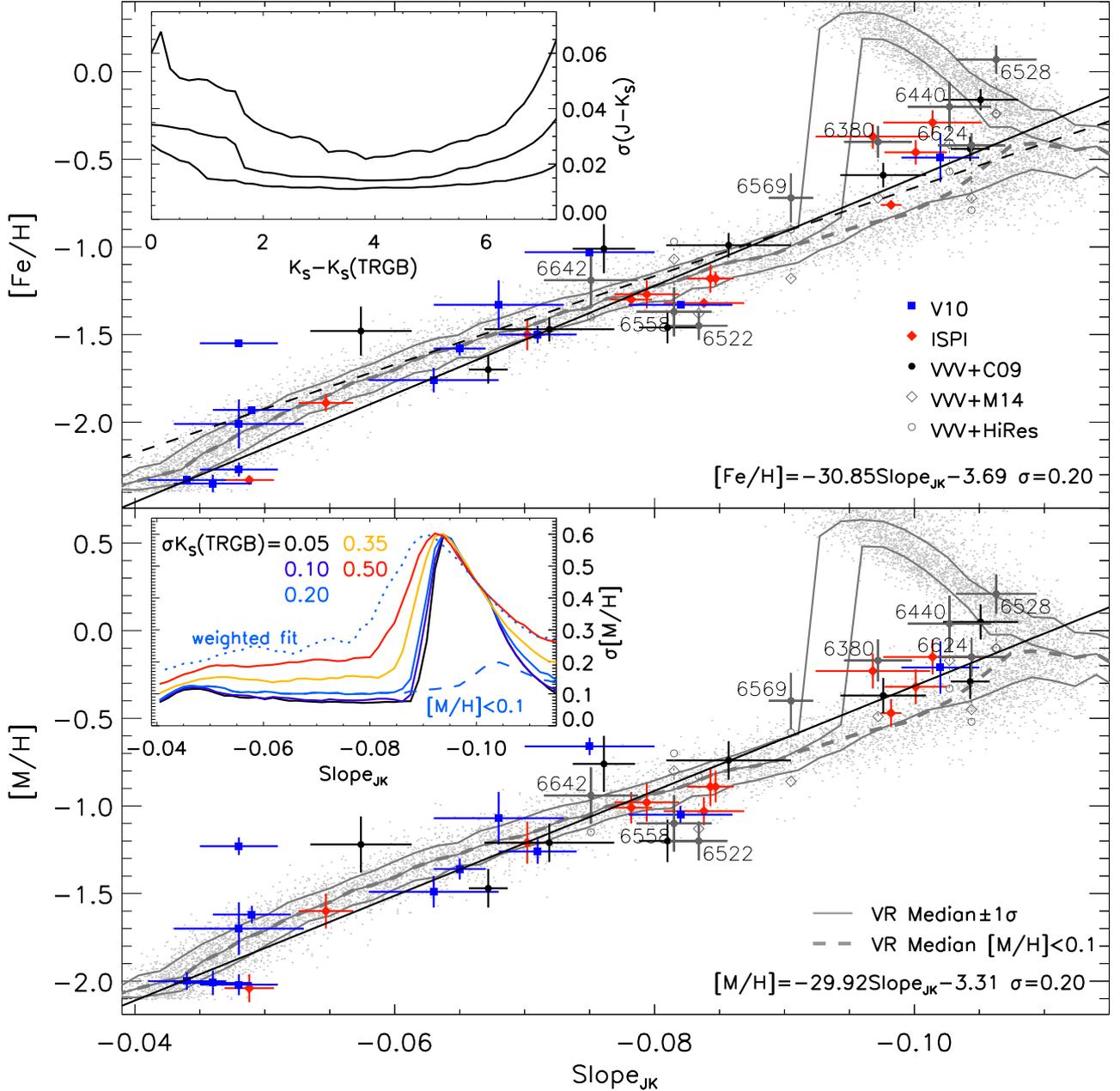}
\caption{Relations between the slope of the red giant branch and cluster metallicity, in terms of
  $\mathit{[Fe/H]}$ (top) and global metallicity $\mathit{[M/H]}$ (bottom).  
  In each plot, values for clusters from the VVV sample are shown as
  filled black circles.  Clusters in the VVV sample which are excluded 
  as calibrators
  due to uncertain metallicities are labelled and plotted in grey rather than
  black , and their \citet{c09} values
  are connected by a dotted line to values corresponding to $\mathit{[Fe/H]}$ from
  \citet{mauro14} (open diamonds) and the HiRes set given in Table
  \ref{fehtab} (open circles).  Additional clusters used as calibrators from
  the V10 sample \citep[][and references therein]{v10} 
  are shown as blue squares, and calibrators
  from \citet{cohenispi} are shown using red circles.  The solid black line
  represents a least squares fit to all calibrators (VVV+V10+ISPI) weighted
  using the uncertainties in metallicity, and the
  resulting best-fitting equation is given in the bottom right corner of each panel.
  The dashed black 
  line represents the relation of \citet{v04obs} transformed to the
  \citet{c09} metallicity scale.  The curved grey lines represent the median
  and $\pm$1$\sigma$ values predicted from monte carlo simulations using 
  Victoria-Regina evolutionary models (see text for details), and 
  the individual simulation results are shown as light grey points.} 
\label{slopefig}
\end{figure*}

In order to assess the impact of the fitting method, photometric errors, and 
observational uncertainty in the TRGB magnitude, we have performed 
an additional series of simulations to examine systematic errors on
the measurement of the RGB slope.  We have generated 1000
synthetic RGBs distributed evenly over the metallicity range
-2.5$<$$\mathit{[Fe/H]}$$<$0.5 using 12 Gyr $\alpha$-enhanced ($[\alpha/Fe]$=0.4)
Victoria-Regina isochrones \citep{vandenberg14} as these models reasonably
reproduce the upper RGB morphology of GGCs in the near-IR (see
\citealt{cohenispi} for details).  For each iteration, the total number of RGB
stars and a value of the LF exponent $B$ were randomly drawn from a Gaussian
distribution with the observed mean and standard deviation.  
Next, all stars were offset using photometric
errors randomly drawn from Gaussian distributions with standard deviations equal
to the observed median photometric error as a function of magnitude 
below the TRGB  
across all of our target clusters, 
shown in the inset in the upper panel of Fig.~\ref{slopefig}.  
Finally, a random measurement uncertainty of
0.2 mag on the TRGB magnitude 
is added (we explore the choice of this value below), 
before measuring the RGB slope 
identically as for the target clusters.  
In the main panels of Fig.~\ref{slopefig},  
the median and $\pm$1$\sigma$ 
standard deviation of the slopes from the synthetic
RGBs (measured in 50 evenly spaced bins) are shown as curved grey lines.  

Because RGBs are
typically described in near-IR CMDs using 
low order polynomials \citep[e.g.][]{v04obs} as we have done, 
it is already known that the
use of a line to fit the upper RGB is a first-order approximation
\citep{ferraro00}, and models predict this.  
As the upper RGB becomes increasingly negative in slope 
(less vertical in the CMD) at higher
metallicities, its curvature increases as well.  For this reason, linear
fits to the upper RGB become increasingly degenerate at higher
($[M/H]\gtrsim$$-$0.5) metallicities.  However, the simulations show that this
effect depends entirely on the maximum assumed 
metallicity limit at the metal-rich end.  In other words, the relation between
slope and metallicity is no longer monotonic at the metal-rich end (or
equivalently $slope_{JK}$$<$-0.09), but this is due entirely 
to the inclusion of 
metallicities ranging well above solar in our simulations.  To illustrate this
conclusion, we have re-evaluated the median and 1$\sigma$ values of metallicity
as a function of slope from the simulations, but excluding all iterations with
an input $[M/H]$$>$-0.1.  The resulting median values are shown as a dotted
grey line in the main panels of Fig.~\ref{slopefig}, 
illustrating that according to the models, the slope can remain an
effective metallicity indicator at high (near-solar) metallicities 
only if super-solar metallicities can be excluded \textit{a priori}.  However, 
the data do not show any evidence for such a degeneracy, and in fact
the more metal-rich calibrating clusters ($slope_{JK}$$\leq$-0.09) show 
rms deviations from our linear fit of only 0.15 and 0.12 dex 
versus $[Fe/H]$ and $[M/H]$ respectively, as compared to an rms of 0.22 dex
versus both $[Fe/H]$ and $[M/H]$ at lower metallicities.

We have also performed sets of simulations to explore two
additional sources of uncertainty in the slope measurements.  The first of
these is the observational uncertainty of the TRGB magnitude.  We have 
reperformed the 1000-iteration simulation several times, assuming a different 
observational uncertainty on the location of the TRGB ranging over
$\sigma$$K_{S}(TRGB)$=(0.05,0.1,0.2,0.35,0.5) mag in each simulation.
In the inset in the lower panel of Fig.~\ref{slopefig}, we plot the median
scatter (e.g.~standard deviation) in metallicity as a function of slope, colour
coded by the input (Gaussian) uncertainty in the TRGB magnitude.  These 
simulations reveal that an uncertainty of 
up to $\sim$0.2 mag in the TRGB location does
not significantly impact the uncertainty in the inferred  
metallicity above a
lower threshold which is set by the photometric errors and number of available
cluster stars typical of our observations.  
However, if the uncertainty
in the TRGB magnitude increases substantially above
$\sigma$$K_{S}(TRGB)$$\sim$0.2, the scatter in metallicity inferred from a
given slope value is significantly affected, increasing by more than a factor
of two if the TRGB location is not known to better than 0.5 mag.

The second source of systematic uncertainties which can be addressed with such
simulations is the use of photometric errors to weight the stars in the 
relevant CMD region when performing the least-squares fit to measure the
slope.  An additional series of simulations was performed as described above, 
assuming $\sigma$$K_{S}(TRGB)$=0.2 to allow a direct comparison, 
but measuring the slopes by performing a weighted rather than unweighted least
squares fit.  The results of this set of simulations are shown as a dotted
blue line in the inset in the lower panel of Fig.~\ref{slopefig}, revealing
that the uncertainty of the inferred metallicities increases by a factor
of more than two when a
weighting scheme is used.  This result is 
specific to the distribution of photometric error versus
magnitude for our target clusters, and the cause is illustrated in the inset 
in the upper panel of Fig.~\ref{slopefig}.   
The necessity of 2MASS
photometry due to saturation in our VVV PSF photometry 
close to the TRGB causes the photometric errors to increase at the bright end
of the magnitude range where the slope is measured.    
The downweighting of these stars in a photometric-error-weighted fit 
combines with their relative
sparseness at the brighter, more poorly populated end of the RGB LF to result
in larger scatter in the measured slopes.  For this reason, we have employed
unweighted least-squares fits when measuring the RGB slopes of our target
clusters (also, the use of an unweighted fit is presumably more
consistent with previous studies as they do not mention a weighting scheme).

Lastly, we address the influence of uncertainties in the decontamination
procedure on our measured slope values.  The decontamination procedure gives a
formal 1$\sigma$ uncertainty on the number of stars in the cluster region which are
probable members, based on both photometric errors as well as Poissonian
uncertainties on the number of stars in the cluster and comparison regions.
By combining this quantity with the membership probability as a function of
location in the cluster CMD, we can check whether any of the stars used in the
calculation of the RGB slope have membership probabilities placing them within
the 1$\sigma$ error margin of $N_{clus}$ 
(given in Table \ref{tab:cleanparams}).  
We find that of the stars used to
calculate the RGB slope, in all target clusters  
less than 5\% of them have membership probabilities placing
them within this 1$\sigma$ error margin, and we have verified
that the inclusion or exclusion of these stars does not affect the measured 
slopes beyond their uncertainties.  Furthermore, 
in half of our target clusters, none of the stars used to measure the slope
are 1$\sigma$ non-members.

\subsection{HB-Bump Magnitude Difference ($\Delta$$m^{HB}_{RGBB}$)}
\label{bumphbsect}
The magnitude difference between the 
HB and the RGBB is another
distance- and reddening-independent metallicity indicator which has not yet
been explored in the near-IR. 
At optical wavelengths, a linear relation was found in the $I$ band 
between the magnitude of the 
(CMD-selected) red HB and the RGBB magnitude \citep{natafbump}.  
Meanwhile several other studies 
\citep{cs97,atabump,zoccalibump,riellobump,diceccobump} have found 
a somewhat non-linear relation between cluster metallicity and the magnitude
difference between the RGBB and the ZAHB
in the $V$ band in accord with predictions of evolutionary models.  However,
among these studies, several different metallicity scales and methodologies
for quantifying the ZAHB magnitude were employed.

We perform fits to the magnitude difference between the HB and the RGBB,
denoted $\Delta$$m^{HB}_{RGBB}$, 
 as a function of metallicity in all three near-infrared $JHK_{S}$ 
bandpasses.  Unlike some of the aforementioned optical studies, the HB
magnitude which we employ corresponds to the peak of the
observed LF, with an uncertainty ascertained through bootstrap resampling.  
Importantly, this procedure allows for 
non-Gaussian HB magnitude distributions, which is particularly relevant in the
near-IR, where the HB may only be truly horizontal at near-solar
metallicities.  The uncertainty of $\Delta$$m^{HB}_{RGBB}$ is calculated as the
quadrature sum of the reported uncertainties on the HB and RGBB magnitude, and 
we conservatively assume an uncertainty of 0.1 mag on the HB magnitude for 
clusters in \citet{v04abs} and \citet{v10} without HB magnitude uncertainties.

\begin{figure*}
\includegraphics[width=0.97\textwidth]{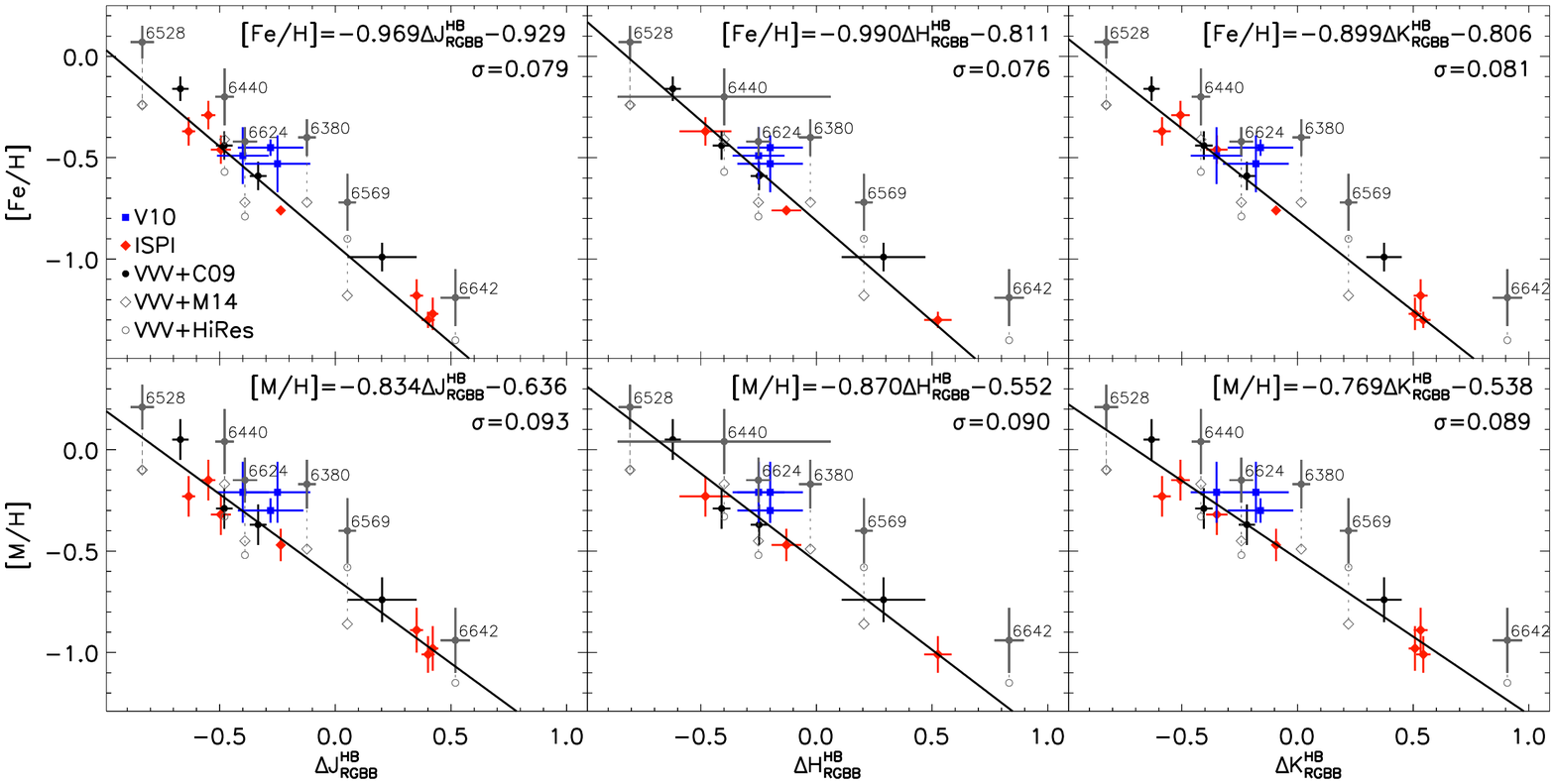}
\caption{Linear fits in all three $JHK_{S}$ filters to 
metallicity as a function of the magnitude difference
  between the RGBB and the HB.  Symbols are as
in Fig.~\ref{slopefig}.}
\label{bumphbfig}
\end{figure*}

In Fig.~\ref{bumphbfig} we
show linear fits of both $\mathit{[Fe/H]}$ and $\mathit{[M/H]}$ as a function of
$\Delta$$m^{HB}_{RGBB}$ (in this case using uncertainties in both axes to weight the
fits cf.~\citealt{cohenispi}), 
with the resulting coefficients given in each panel of
Fig.~\ref{bumphbfig} and summarized in Table \ref{relationtab}. 
As only the relatively metal-rich
($[M/H]\gtrsim$-1.1) clusters in our sample 
show a detectable peak in their LF resulting
from the HB, the $\Delta$$m^{HB}_{RGBB}$ diagnostic is 
only applicable at these higher metallicities, 
but nevertheless the standard
deviation of the fit residuals is $\leq$0.1 dex in all cases.

However, it should be somewhat surprising that the magnitude difference between the
RGBB and the HB is well fit by a linear relation since  
current
observational and theoretical evidence implies that the luminosity of neither
the RGBB nor the HB is strictly a linear function of metallicity.  
On the observational
side, \citet{v04abs} and \citet{cohenispi} 
used a quadratic relation to fit the near-IR bump 
luminosity versus metallicity, while from a theoretical perspective, the
models of \citet{sg02} suggest that the $K_{S}$ luminosity of the HB
is a non-linear function of metallicity even at fixed age.  
Along similar lines, both the RGBB and HB magnitudes are
predicted to depend on second, and likely third
parameters in addition to metallicity.  
The \citet{sg02} models predict that
the $K_{S}$ luminosity of the HB depends on age as well as 
metallicity (at a level of $\leq$0.05 mag/Gyr for typical GGC ages), although 
deep IR photometry of optically well-studied clusters is
needed to confirm this predictions.  Regarding the RGBB, \citet{47tucir}
point out that at least at the metallicity of 47 Tuc, $\alpha$-enhanced BaSTI 
models \citep{basti} predict that the near-IR luminosity of
the RGBB is affected by changes in $[\alpha/Fe]$ (at a level of
$\Delta$$K_{S}$=0.06 mag for $\Delta$$[\alpha/Fe]$=0.4), 
and to a somewhat lesser extent by age (although in
this case models predict that the RGBB and HB change in the same direction). 
The data which we employ
are insufficient to confirm or deny whether a higher
order fit to the $\Delta$$m^{HB}_{RGBB}$-metallicity relation is
appropriate or not, and more secure ages and 
spectroscopic abundances for bulge GGCs could improve the situation.  However,
the current typical uncertainties of $\sim$0.1 dex on global metallicity and 
$\sim$0.05 mag on $\Delta$$m^{HB}_{RGBB}$ would not permit the detection of a subtle 
non-linearity in the relation, and indeed the relations between metallicity and near-IR
RGBB and HB magnitude from \citet{cohenispi}, along with their uncertainties, imply that
in the higher-metallicity regime explored here ($[M/H]$$<$1), a relation between
metallicity and $\Delta$$m^{HB}_{RGBB}$ is expected to be linear at the $\sim$0.1 dex level.

\subsection{RGB bump-tip Magnitude Difference ($\Delta$$m^{RGBB}_{TRGB}$)}
\label{bumptipsect}
\begin{figure*}
\includegraphics[width=0.97\textwidth]{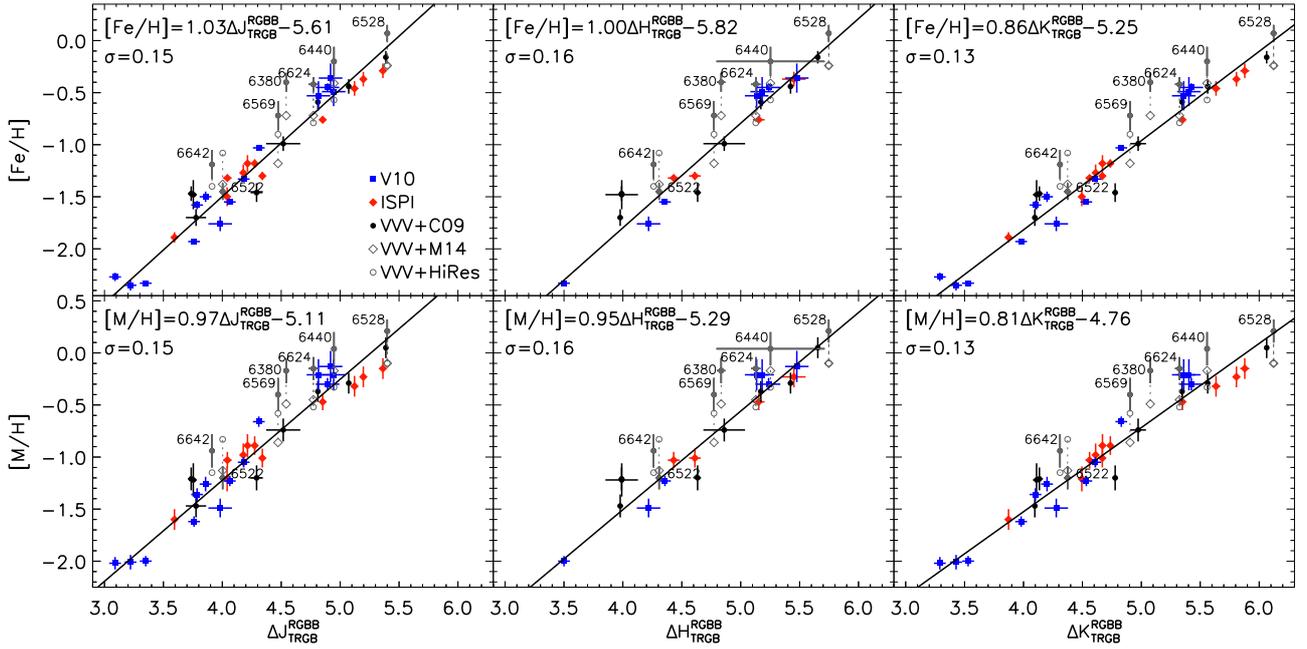}
\caption{Linear fits in all three $JHK_{S}$ filters to
metallicity as a function of the magnitude difference
between the RGBB and the TRGB.  Note the difference in
axis scales as compared to Fig.~\ref{bumphbfig}.  
Symbols are as in Fig.~\ref{slopefig}, but the horizontal error bars
represent only uncertainties in the RGBB magnitude and are not used
in the fits.}
\label{bumptipfig}
\end{figure*}

The parameter $\Delta$$m^{RGBB}_{TRGB}$ is another potential metallicity indicator
which we explore in the infrared for the first time.  
Additionally, given that the RGBB magnitude can generally be measured with
much greater precision than the TRGB magnitude (see Sect.~\ref{tipsect},
Appendix \ref{tipdetailsect} and
Table \ref{obstab}), we also explore the precision of the $\Delta$$m^{RGBB}_{TRGB}$
vs. metallicity relations as a vehicle to empirically quantify the uncertainty
of the TRGB magnitude.
Relations between $\Delta$$m^{RGBB}_{TRGB}$ and metallicity are shown in
Fig.~\ref{bumptipfig}, and as the uncertainty in the TRGB location dominates
the uncertainty in that of the RGBB, we weight our least squares fits only by
the (y-axis) uncertainty in metallicity.   
If we invert the best-fit relations and calculate
the rms residuals of a fit with respect to $\Delta$$m^{RGBB}_{TRGB}$ 
rather than metallicity,
we obtain values of 0.14$<$$\sigma$$\Delta$$m^{RGBB}_{TRGB}$$<$0.17 mag.  As the
uncertainty in the RGBB location generally contributes negligibly to this 
quantity (The VVV and ISPI samples have median uncertainties of $<$0.03 mag on
$m_{RGBB}$), it would appear that the median uncertainty on the location of
the TRGB is
$<$0.2 mag in the near-IR, somewhat smaller than the typical values obtained by
\citet[][see their table 3]{v04abs} based on the  prescription employed by
\citet{ferraro99}: The median of their reported TRGB uncertainties is
(0.25,0.23,0.26) mag in ($J,H,K_{S}$).  However, 
the discussion in Appendix \ref{tipdetailsect} is presented to highlight 
the complexity of attempting to measure the TRGB location in GGCs, and when
viewed on a case-by-case basis, ambiguities in the TRGB location are
often larger than a naive extrapolation of the residuals of our linear 
fits in Fig.~\ref{bumptipfig} would imply.

For convenience, the coefficients of the 
linear fits shown in Figs.~\ref{slopefig}-\ref{bumptipfig}, 
along with their uncertainties and the rms
residuals of the fits, are summarized 
in Table \ref{relationtab}.

\begin{table}
\centering
\caption{Linear Coefficients for Photometric Metallicity Indicators: $y=A+Bx$}
\begin{tabular}{lcccc}
\hline
 $x$ & $y$ & $A$ & $B$ & RMS \\
\hline
$slope_{JK}$ & $\mathit{[Fe/H]}$ & -3.69$\pm$0.06 & -30.85$\pm$0.80 & 0.20 \\
$\Delta$$J^{HB}_{RGBB}$ & $\mathit{[Fe/H]}$ & -0.929$\pm$0.020 & -0.969$\pm$0.050 &
0.079 \\
$\Delta$$H^{HB}_{RGBB}$ & $\mathit{[Fe/H]}$ & -0.811$\pm$0.034 & -0.990$\pm$0.080 &
0.076 \\
$\Delta$$K^{HB}_{RGBB}$ & $\mathit{[Fe/H]}$ & -0.806$\pm$0.018 & -0.899$\pm$0.047 &
0.081 \\
$\Delta$$J^{RGBB}_{TRGB}$ & $\mathit{[Fe/H]}$ & -5.61$\pm$0.27 & 1.03$\pm$0.06 &
0.15 \\
$\Delta$$H^{RGBB}_{TRGB}$ & $\mathit{[Fe/H]}$ & -5.82$\pm$0.40 & 1.00$\pm$0.08 &
0.16 \\
$\Delta$$K^{RGBB}_{TRGB}$ & $\mathit{[Fe/H]}$ & -5.25$\pm$0.22 & 0.86$\pm$0.05 &
0.11 \\
$slope_{JK}$ & $\mathit{[M/H]}$ & -3.31$\pm$0.09 & -29.92$\pm$1.10 & 0.20 \\
$\Delta$$J^{HB}_{RGBB}$ & $\mathit{[M/H]}$ & -0.636$\pm$0.034 & -0.834$\pm$0.077 &
0.093 \\
$\Delta$$H^{HB}_{RGBB}$ & $\mathit{[M/H]}$ & -0.552$\pm$0.045 & -0.870$\pm$0.114 &
0.090 \\
$\Delta$$K^{HB}_{RGBB}$ & $\mathit{[M/H]}$ & -0.538$\pm$0.031 & -0.769$\pm$0.070 &
0.089 \\
$\Delta$$J^{RGBB}_{TRGB}$ & $\mathit{[M/H]}$ & -5.11$\pm$0.27 & 0.97$\pm$0.06 &
0.15 \\
$\Delta$$H^{RGBB}_{TRGB}$ & $\mathit{[M/H]}$ & -5.29$\pm$0.39 & 0.95$\pm$0.08 &
0.16 \\
$\Delta$$K^{RGBB}_{TRGB}$ & $\mathit{[M/H]}$ & -4.76$\pm$0.22 & 0.81$\pm$0.05 &
0.13 \\
\hline
\end{tabular}
\label{relationtab}
\end{table}

\section{Discussion}
\label{discusssect}

Our results shown in Figs.~\ref{slopefig}, \ref{bumphbfig} and \ref{bumptipfig}
suggest that the three photometric metallicity indicators $slope_{JK}$, $\Delta$$m^{HB}_{RGBB}$ and $\Delta$$m^{RGBB}_{TRGB}$ each have their respective advantages in different metallicity regimes.  At relatively high ($[M/H]$$\gtrsim$-1) metallicities, $\Delta$$m^{HB}_{RGBB}$ yields the best overall precision with an rms deviation of $<$0.1 dex from our linear fit in all three $JHK_{S}$ filters, although both  
$slope_{JK}$ and $\Delta$$m^{RGBB}_{TRGB}$ do nearly as well, with rms deviations of $\sim$0.15 dex.  However, moving to lower metallicities, $\Delta$$m^{HB}_{RGBB}$ becomes difficult to apply for two reasons.   
First, clusters which are more metal-rich tend to have HB
magnitudes which can be more reliably measured in the near-IR due to the
increased horizontality of the HB in near-IR CMDs.  For example, 
a peak in the cluster LF corresponding to the
HB could not be reliably detected for clusters with
$\mathit{[Fe/H]}$\citep{c09}$<$-1.2, 
similar to the results of \citet{cohenispi}.  
Second, the RGB bump becomes less 
prominent with decreasing metallicity \citep{natafbump}, also hindering the
use of $\Delta$$m^{RGBB}_{TRGB}$.  Therefore, despite its somewhat poorer rms deviation of $\sim$0.2 dex, $slope_{JK}$ may be the best option at lower metallicities, particularly for relatively sparse stellar populations where either the RGBB and/or the TRGB location cannot be reliably measured.

\begin{figure*}
\includegraphics[width=0.99\textwidth]{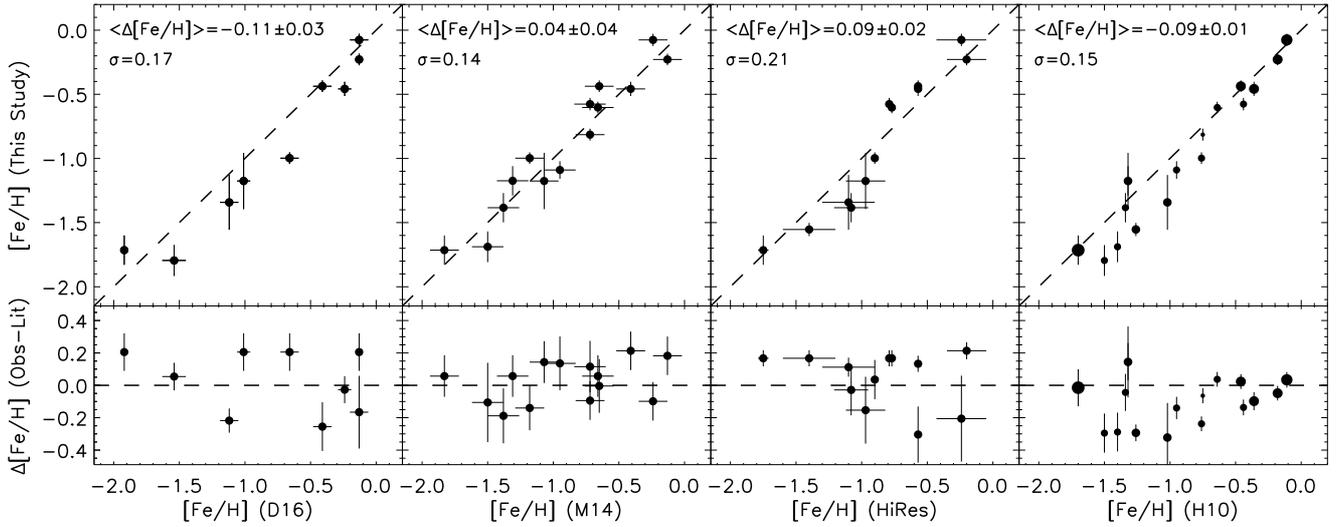}
\caption{Comparison between our photometric $\mathit{[Fe/H]}$ estimates given in
  Table \ref{photcaltab} and $\mathit{[Fe/H]}$ values from (left to right)
  \citet{brunolores}, \citet{mauro14}, the HiRes values in Table \ref{fehtab} 
and the \citet{h96} catalog.  In each plot, the dashed line indicates
equality.  Because \citet{h96} give weights rather than formal uncertainties
on their $\mathit{[Fe/H]}$ values, the size of the plotting symbol is
proportional to the weight given to the $\mathit{[Fe/H]}$ value for each
cluster in the rightmost panel by \citet{h96}.} 
\label{photcalfig}
\end{figure*}

We can apply the calibrations listed in Table \ref{relationtab} to obtain
purely photometric metallicity estimates for our target clusters.  For each
cluster, there are a total of seven calibrations available to calculate
$\mathit{[Fe/H]}$ or $\mathit{[M/H]}$, and we exclude
those for which data are not available in individual cases (i.e. the HB
magnitudes for metal-poor clusters).  The resulting 
mean photometric $\mathit{[Fe/H]}$ and $\mathit{[M/H]}$, weighted by 
the inverse quadrature sum 
of the observational uncertainty and the calibration rms, are given in Table
\ref{photcaltab} along with the number of relations from Table
\ref{relationtab} available.  These photometric metallicity estimates are
compared in Fig.~\ref{photcalfig} with $\mathit{[Fe/H]}$ values from
\citet{brunolores},\citet{mauro14}, the HiRes spectroscopic values in Table
\ref{fehtab}, and the \citet{h96} catalog.

\begin{table}
\centering
\caption{Photometric Metallicity Estimates for Target Clusters}
\begin{tabular}{lccc}
\hline
 Cluster & $\mathit{[Fe/H]}$ & $\mathit{[M/H]}$ & N(relations) \\
\hline
NGC6380 & -0.81$\pm$0.05 & -0.55$\pm$0.05 & 7 \\
NGC6401 & -1.34$\pm$0.21 & -1.03$\pm$0.21 & 1 \\
NGC6440 & -0.46$\pm$0.06 & -0.23$\pm$0.06 & 7 \\
NGC6441 & -0.44$\pm$0.05 & -0.21$\pm$0.05 & 7 \\
NGC6453 & -1.80$\pm$0.12 & -1.49$\pm$0.12 & 4 \\
NGC6522 & -1.38$\pm$0.11 & -1.10$\pm$0.11 & 4 \\
NGC6528 & -0.08$\pm$0.05 & 0.10$\pm$0.05 & 7 \\
NGC6544 & -1.69$\pm$0.12 & -1.39$\pm$0.12 & 4 \\
NGC6553 & -0.23$\pm$0.05 & -0.03$\pm$0.05 & 7 \\
NGC6558 & -1.18$\pm$0.22 & -0.87$\pm$0.22 & 1 \\
NGC6569 & -1.00$\pm$0.05 & -0.71$\pm$0.05 & 7 \\
NGC6624 & -0.58$\pm$0.05 & -0.34$\pm$0.05 & 7 \\
M28 & -1.17$\pm$0.11 & -0.90$\pm$0.11 & 4 \\
M69 & -0.60$\pm$0.05 & -0.37$\pm$0.05 & 7 \\
NGC6638 & -1.09$\pm$0.07 & -0.79$\pm$0.07 & 7 \\
NGC6642 & -1.55$\pm$0.05 & -1.20$\pm$0.05 & 7 \\
M22 & -1.71$\pm$0.11 & -1.41$\pm$0.11 & 4 \\
\hline
\end{tabular}
\label{photcaltab}
\end{table}

In general, our linear fits in 
Figs.~\ref{slopefig}-\ref{bumptipfig} favor recent
spectroscopic metallicities over those listed in the compilation of 
\citet{c09} for metal-rich clusters 
(NGC 6380, 6440, 6528, 6569 and to a lesser extent NGC 6624). 
For the six target clusters with $\mathit{[Fe/H]}$$\lesssim$-0.7, 
the agreement with the \citet{h96} catalog is particularly good, with a 
mean offset of -0.03$\pm$0.02 dex.
For NGC 6528, arguably the most metal-rich GGC, our calibrations 
give $\mathit{[Fe/H]}$ between the
the lower values reported by \citet{zoccali04},  
\citet{origlia05} and \citet{sobeck06} 
and the super-solar value of \citet{carretta01}, 
in good agreement with the low-resolution spectra of \citet{dias14}.

In a global sense, our photometric metallicity calibrations agree best
with the CaII triplet values of \citet{mauro14} compared to the other 
sets of literature metallicity values.  
For example, our photometric metallicities imply a decrease in 
$\mathit{[Fe/H]}$ of $\sim$0.3-0.4 dex for NGC 6380 and NGC 6569 compared to
their \citet{c09} values.  For NGC 6380, this is also in agreement with the
\citet{h96} catalog, while NGC 6569 is 
a significant outlier in the CaII triplet calibration of \citet{mauro14}, who 
found $\mathit{[Fe/H]}$=-1.18$\pm$0.11, in better ($\sim$2$\sigma$) accord
with our photometric metallicities than any spectroscopic results.
For NGC 6401 and NGC 6558, our photometric values are $\sim$0.2 dex lower than
those measured from spectroscopy.  However, because these clusters lack a
detectable RGBB or red HB, our photometric metallicity estimate is based only
on the RGB slope, so the uncertainties are relatively large.  Furthermore, since
these clusters are relatively sparse and projected on the Galactic bulge, 
their TRGB magnitudes remain uncertain at
the $\gtrsim$0.3 mag level (see Appendix \ref{tipdetailsect}).  This implies
yet a  
larger corresponding uncertainty of the RGB slope and hence the photometric
metallicity should our chosen TRGB candidate be proven incorrect.
For the remainder of metal-intermediate blue HB clusters in the VVV sample, 
we also find metallicities $>$0.2-0.3 dex lower than those of \citet{c09}. 
For NGC 6544, the \citet{c09}
value is supported by \citet{mauro14}, although the discrepancy between their 
result and ours is only marginally significant in light of the large
uncertainties.  Meanwhile, our photometric $\mathit{[Fe/H]}$ value for 
NGC 6453 rests fairly heavily on the assumed TRGB magnitude, 
and in fact our photometric $\mathit{[M/H]}$ value is in good
agreement with \citet{brunolores} if this cluster has a
relatively low level of $\alpha$-enhancement as suggested by their fits 
to synthetic spectra.  Lastly, for NGC 6642, the value given
by our photometric calibrations
agrees with \citet{minniti95} to within the uncertainties, 
and a value as
low as $\mathit{[Fe/H]}$=-1.8 was suggested by \citet{balbinot} based on 
isochrone fitting to space-based optical photometry.   

Our results underscore the need for high-resolution multi-object spectroscopy 
of poorly studied bulge GGCs.  The tendency of the \citet{c09} compilation to
overestimate the $\mathit{[Fe/H]}$ of metal-rich bulge clusters could be simply an
artefact of 
high field star densities in the original integrated light studies
compiled by \citet{c09} and/or the use of a super-solar metallicity for NGC 6528
to convert previous metallicity scales to their UVES scale.  In either case,
the GGC metallicity scale at high metallicities remains poorly constrained,
and detailed spectroscopic analyses of large samples of cluster stars (for
example, to assess contamination by AGB members) are crucial for accurate
and self-consistent determinations of $\mathit{[Fe/H]}$ as well as 
$[\alpha/Fe]$.  This
would be a valuable step towards testing GGC evolutionary models at near-solar
metallicities and improving our relations to allow a deeper understanding of
distant, composite and/or heavily extincted stellar populations. 

\section*{Acknowledgements}

It is a pleasure to thank Jim Emerson for discussions regarding the
calibration of VISTA photometry to the 2MASS photometric system, and
the anonymous referee for their insightful comments, which enhanced the
quality and presentation of this study.  
REC is grateful for financial support from Fondo GEMINI-CONICYT 32140007, 
and FM is thankful for financial support from FONDECYT for project 3140177.
DG, FM and REC also acknowledge financial support from the Chilean BASAL
Centro de Excelencia en Astrofisica y Technologias Afines (CATA) grant
PFB-06/2007.  CMB acknowledges financial support from FONDECYT through regular 
project 1150060.  This publication makes use of data products from the
Two Micron All Sky Survey, which is a joint project of the University of
Massachusetts and the Infrared Processing and Analysis Center/California
Institute of Technology, funded by the National Aeronautics and Space
Administration and the National Science Foundation.  This publication is based
largely on observations collected at the European Organisation for
Astronomical Research in the Southern Hemisphere, Chile, under ESO program
179.B-2002 (VVV survey).

\appendix
\section{Details of the TRGB Identification}
\label{tipdetailsect}
Here we give the details concerning the choice of the brightest RGB member
star in each cluster.  Care has been
taken to use published proper
motion, photometric variability, radial velocity and/or chemical abundance studies as well as
additional photometry from the literature where available in order to
assess the likelihood that a CMD-selected TRGB candidate is a cluster RGB
member.  As this still leaves the choice of TRGB star ambiguous in some cases, 
radial location in the cluster is also employed to judge the membership of
TRGB candidates,  
so in many
cases we refer to the radial distance from the centre of the cluster 
as a fraction of the core radius
($R_{c}$) or half-light radius ($R_{hl}$) from the \citet{h96}
catalog\footnote{The structural parameters from \citet{h96} may not be
  reliable in some cases, as discussed by \citet{mvdm}}.  
Clusters are listed by the source of their
photometry, with the VVV clusters from the present study first, followed by those from 
\citet{cohenispi}, and
finally those from \citet{v04abs} and \citet{v10} absent from the two more
recent studies.  The TRGB candidate which we have chosen in each cluster is
given in Table \ref{tipstartab}, along with its position, photometry, 2MASS
ID, and the source of the given position and photometry.  Although all TRGB
candidates could be reliably matched to 2MASS counterparts, where possible we
have employed the photometry from the given source (corrected for differential
reddening in the case of the VVV clusters) in place of 2MASS due to the 
improved spatial resolution of the source catalog over the 2MASS PSC, as well
as photometric quality flags in 2MASS warning of low quality photometry.
Conversely, in some cases available near-IR catalogs saturate below the TRGB,
requiring the use of photometry from 2MASS. 

\begin{table*}
\centering
\caption{Selected TRGB Candidates}
\begin{tabular}{lrrrrrcl}
\hline
 Cluster & RA(J2000) & Dec(J2000) & $J(TRGB)$ & $H(TRGB)$ & $K_{S}(TRGB)$ &
 2MASS ID & Source \\
\hline
 NGC6380 & 263.600525 & -39.064743 &  10.526 &   9.331 &   8.770 & J17342412-3903530 & 2MASS \\
 NGC6401 & 264.650427 & -23.907622 &  10.436 &   9.348 &   9.040 & J17383610-2354275 &   V10 \\
 NGC6440 & 267.226342 & -20.367685 &  10.214 &   8.977 &   8.459 & J14485434-2022034 &   V10 \\
 NGC6441 & 267.558868 & -37.065826 &  10.532 &   9.470 &   9.188 & J17501414-3703569 &   V10 \\
 NGC6453 & 267.715425 & -34.610149 &  10.622 &   9.673 &   9.402 & J17505170-3436365 &   V10 \\
 NGC6522 & 270.871602 & -30.046986 &   9.863 &   8.882 &   8.649 & J18032918-3002491 &   V10 \\
 NGC6528 & 271.182471 & -30.047783 &   9.338 &   8.265 &   7.739 & J18044378-3002523 &   V10 \\
 NGC6544 & 271.846548 & -24.976402 &   7.715 &   6.713 &   6.364 & J18072317-2458350 & 2MASS \\
 NGC6553 & 272.325166 & -25.911547 &   8.499 &   7.415 &   6.812 & J18091804-2254415 &   V10 \\
 NGC6558 & 272.573617 & -31.760925 &  10.079 &   9.065 &   8.822 & J18101766-3145393 & 2MASS \\
 NGC6569 & 273.410993 & -31.835494 &  10.471 &   9.479 &   9.192 & J18133863-3150077 &   V10 \\
 NGC6624 & 275.943807 & -30.317759 &   9.536 &   8.534 &   8.234 & J18234651-3019039 & 2MASS \\
     M28 & 276.174360 & -24.884188 &   8.825 &   7.852 &   7.548 & J18244184-2453030 & 2MASS \\
     M69 & 275.885411 & -32.293312 &   9.574 &   8.637 &   8.358 &  J1831249-3217359 & 2MASS \\
 NGC6638 & 277.732604 & -25.500245 &   9.936 &   8.946 &   8.682 & J18305581-2530007 &   V10 \\
 NGC6642 & 277.970396 & -23.476840 &  10.023 &   9.031 &   8.830 & J18315289-2328365 &   V10 \\
     M22 & 277.062931 & -23.915266 &   7.737 &   6.966 &   6.722 & J18361510-2354549 & 2MASS \\
  NGC104 &   6.063092 & -72.076809 &   7.876 &   6.997 &   6.723 & J00241513-7204365 & 2MASS \\
 NGC0288 &  13.171358 & -26.557552 &   9.693 &   8.819 &   8.589 & J00524112-2633271 & 2MASS \\
  NGC362 &  15.821451 & -70.847116 &   9.414 &   8.699 &   8.467 & J01031723-7050496 & 2MASS \\
 NGC1261 &  48.065417 & -55.211288 &  10.819 &  10.025 &   9.808 & J03121569-5512406 &  ISPI \\
 NGC1851 &  78.531782 & -40.040782 &  10.210 &   9.320 &   9.138 & J05140762-4002267 &  ISPI \\
 NGC2808 & 137.987816 & -64.858240 &   9.978 &   9.096 &   8.793 & J09115707-6451296 &  ISPI \\
 NGC4833 & 194.955866 & -70.904366 &   9.323 &   8.493 &   8.260 & J12594940-7054157 & 2MASS \\
 NGC5927 & 231.992006 & -50.656418 &   9.222 &   8.233 &   7.882 & J15275807-5039230 & 2MASS \\
 NGC6304 & 258.638476 & -29.430115 &   9.014 &   7.879 &   7.488 & J17143323-2925484 & 2MASS \\
 NGC6496 & 269.737148 & -44.264393 &   9.661 &   8.709 &   8.381 & J17585691-4415517 & 2MASS \\
 NGC6584 & 274.551755 & -52.170807 &  10.583 &   9.727 &   9.504 & J18181242-5210149 & 2MASS \\
 NGC7099 & 325.089558 & -23.164424 &   9.394 &   8.873 &   8.627 & J21402149-2309518 &  ISPI \\
 NGC5272 & 205.562545 &  28.390408 &   9.842 &   9.198 &   8.900 & J13421508+2823256 &   V04 \\
 NGC5904 & 229.650179 &   2.110380 &   9.068 &   8.210 &   8.041 & J15183604+0206373 & 2MASS \\
 NGC6205 & 250.424819 &  36.447708 &   9.264 &   8.493 &   8.299 & J16414196+3626518 &   V04 \\
 NGC6341 & 259.285217 &  43.137569 &   9.629 &   8.973 &   8.922 & J17170841+4308149 &   V04 \\
 NGC6342 & 260.305511 & -19.572372 &   9.705 &   8.668 &   8.346 & J17211332-1934205 &   V10 \\
 NGC6752 & 287.783875 & -60.031040 &   7.836 &   6.993 &   6.717 & J19110813-6001517 & 2MASS \\
 NGC6273 & 255.655961 & -26.266617 &   9.668 &   8.832 &   8.572 & J17023743-2615599 &   V10 \\
 NGC6316 & 259.148447 & -28.127740 &  10.281 &   9.174 &   8.832 & J17163562-2807398 &   V10 \\
 NGC6355 & 260.994566 & -26.351975 &  10.191 &   9.300 &   8.918 & J17235869-2621071 &   V10 \\
 NGC6388 & 264.037095 & -44.760095 &  10.287 &   9.233 &   8.905 & J17360890-4445363 &   V10 \\
 NGC6539 & 271.212936 &  -7.571320 &  10.084 &   8.915 &   8.470 & J18045110-0734166 &   V10 \\
\hline
\end{tabular}
\label{tipstartab}
\end{table*}

\noindent\textbf{NGC 6380:} The selected tip star has Xflg=2 (signifying
that it is within an extended source) in 2MASS.  There are several 
slightly brighter candidate tip stars present in 2MASS or only in the
\citet{v10} catalog, but all of these have $J-H$ and/or 
$J-K_{S}$ colours somewhat ($\sim$0.1 mag) discrepant from the observed
cluster RGB.  Moving faintward, if our chosen TRGB candidate is not a true RGB star,
selection of the next several fainter candidates with colours compatible
with RGB membership would affect the TRGB 
magnitudes by $\pm$0.1 mag in each of the three filters.

\noindent\textbf{NGC 6401:} Although \citet{v10} select 2MASS J17383033-2352537,
this star may not be a member: After applying differential reddening
corrections, it lies
slightly ($\sim$0.05) blueward of the cluster sequence in $J-H$, and lies at $\sim$1.15$R_{hl}$.  Our selected TRGB candidate
is a much more likely member based both on
differential reddening corrected \citet{v10} photometry 
and distance of only $\sim$9$\arcsec$ ($<$$R_{c}$) from the cluster center.
This represents a faintward revision of 0.2-0.3 mag in the TRGB of this
cluster from the value reported by \citet{v10}, and
we note that the RGB of NGC 6401 is relatively sparse, and 
\citet{chun} chose to refrain from reporting a TRGB magnitude.  
However, if the uncertain RGBB magnitude we report is correct, then
photometric as well as spectroscopic metallicity estimates for this cluster
argue for a significantly brighter ($\gtrsim$0.5 mag) TRGB, which would
also likely move the measured RGB slope value into better accord with clusters
at similar metallicities.  

\noindent\textbf{NGC 6440:} The selected TRGB candidate is the brightest with
$J-K_{S}$ colour consistent with the observed cluster fiducial sequence, but is
not detected in the $H$ band in the \citet{v10} catalog.  This star is
a 2$\sigma$ (0.389$\arcsec$) positional match with 2MASS J17485434-2022034, which
gives a $J-H$ colour consistent with this star being an RGB member, although
2MASS reports Xflg=2.

\noindent\textbf{NGC 6441:} We discard the several brightest CMD-selected
candidates (variables V1, V131, V134, OGLE-BLG-LPV-060919) due to their long periods and 
large pulsational amplitudes.  Our chosen candidate is V23,
for which \citet{layden6441} find evidence of long-term variability but are unable to further constrain pulsational properties.  Furthermore, mean magnitudes from their optical photometry place this star on the cluster RGB.  However, if this candidate should turn out to be an AGB star, the next best candidate is 2MASS J17501619-3702476, for which the differential reddening-corrected \citet{v10} catalog implies a faintward TRGB shift of $<$0.08 mag in all three 
$JHK_{S}$ filters.   

\noindent\textbf{NGC 6453:} The brightest TRGB candidate, also selected by \citet{v10} has a
94\% probability of membership to the open cluster M 9 \citep{diasocs}, so we select the next candidate, for which \citet{diasocs} reports a 0\% membership probability to M 9.

\noindent\textbf{NGC 6522:} The tip star selected by \citet{v10} is V5476 Sgr,
identified as an OGLE Small Amplitude Red Giant \citet[OSARG;][]{oglelpv}, as is the brighter candidate V5471 Sgr. While their
pulsational properties alone do not exclude the possibility of RGB status or cluster membership, in these cases as well as all brighter candidates (V5462 Sgr, V5468 Sgr, V5475 Sgr),  
a comparison with photometry of the surrounding field
from \citet{oglephot} suggests that these are bulge, rather than cluster
giants.  Conversely, our selected TRGB candidate, V5466 Sgr, has mean $V$ and
$I$ magnitudes more consistent with cluster membership.  However, pending
confirmation of membership for any of these variables, the TRGB magnitude may be subject to change by as much as 0.5 mag in all three filters.   

\noindent\textbf{NGC 6528:} The TRGB location for this cluster is also somewhat uncertain. The
TRGB magnitudes we report correspond to differential reddening corrected
photometry of 2MASS J18044378-3002523=OGLE-BLG-LPV-200787. Another brighter candidate,
the OSARG OGLE-BLG-LPV-201338, cannot be excluded from membership based on pulsational properties or photometry, but has proper motions more consistent with bulge than cluster membership \citet{feltzing}.  As in the similarly ambiguous cases above, should our choice of TRGB candidate be proven incorrect, the TRGB magnitudes would be affected by $>$0.2 mag in all three filters.

\noindent\textbf{NGC 6544:} As the cluster and Galactic bulge sequences are 
well separated in the
CMD in this case, 2MASS J18072317-2458350 appears to be a fairly unambiguous
choice. However, this star is 1.48$\arcmin$ ($>$4$R_{hl}$;
\citealt{cohen6544}) from the cluster center, whereas
there is another potential tip star (2MASS J18071937-2459558) 
which is only 0.17 mag fainter in $K_{S}$ and is only 10.3$\arcsec$
($<$$R_{c}$) from the 
cluster center, although its 2MASS photometry may be somewhat unreliable given
its value of Cflg=ddd and Xflg=2.

\noindent\textbf{NGC 6553:} We reject the brightest two CMD-selected TRGB
candidates in our decontaminated catalogs, V4 and V5.  V4 is a Mira variable,
and both have colours inconsistent with the location of the cluster RGB (at the
0.1 mag level in $(J-H)$).  We select the same TRGB candidate as \citet{v10}, which lies well inside the cluster core radius.
 
\noindent\textbf{NGC 6558:} This is yet another case where the TRGB determination is
particularly uncertain, and where \citet{chun} did not report the TRGB
magnitude from their near-IR photometry.
The brightest potential candidate is V2, which may in fact be constant, and \citet{samus} noted that this star may be affected by blending.  The next two brightest candidates lack any discriminating membership information aside from our CMDs, and our chosen TRGB candidate has a 2MASS position which is a $\sim$1.9 arcsec match to star NGC 6558\_8 from \citet{dias14}.  If a true match, this is a spectroscopically confirmed member (also see \citealt{zoccalibulge}), and lies much closer to the cluster center (0.2$\arcmin$$\sim$0.1$R_{hl}$) 
than the two brighter candidates ($>$1.3$\arcmin$).  However, if any of the brighter candidates is confirmed as an RGB member, the TRGB magnitude of this cluster could move brightward by $>$0.3 mag.  
  
\noindent\textbf{NGC 6569:}We have excluded the long period variables V3 and
V21.  Although the former has no additional pulsational properties listed,
2MASS warns of low-quality photometry.
We have also excluded 2MASS J18133939-3149209 in light of its blue colour in
the \citet{v10} catalog (this star also has low quality photometry in 2MASS).
However, if either this star or V3 are confirmed as RGB members, the TRGB magnitude would move brightward by $>$0.2 mag. 

\noindent\textbf{NGC 6624:}The only TRGB candidates brighter than our
selection have colours from both 2MASS and the \citet{v10} catalog inconsistent
with the location of the cluster RGB.  However, as our candidate lies
2.9$\arcmin$ ($\sim$3.5$R_{hl}$) from the cluster center, if either of these
are revealed to be RGB members, the TRGB magnitude could move brightward by
$>$0.2 mag.  Conversely, if none of these stars, including our selected TRGB candidate, are members, the TRGB magnitude would move faintward from the values we report by $>$0.13 mag in all three filters.

\noindent\textbf{NGC 6626 (M28):} The brightest three TRGB candidates in this cluster all lie to the blue side of the cluster RGB in $(J-H)$ colour.  Of these, one is NV8, suggested by \citet{prieto} to be a Type II Cepheid based on its period and light curve.  Another, V10, has a membership probability of 90\% \citep{pmm28} and an amplitude of $A_{V}$=0.6, and cannot be definitively excluded as an RGB star.  Meanwhile, the brightest of the three in $K_{S}$ is not a known variable and is absent from the proper motion study of \citet{pmm28}, but sits $\sim$0.08 mag blueward of the cluster RGB.  As 2MASS indicates excellent photometric quality for this star, we exclude it as a candidate based on its blue colour.  Our selected candidate is the brightest star with colours in excellent agreement with the observed cluster RGB, and is star 2-56 in \citet{pmm28}, who give a membership probability of 92\%.  In addition, its $K_{S}$ magnitude of 7.548$\pm$0.024 is in reasonable agreement with the
RGB tip location of $K_{S}$(TRGB)=7.45$\pm$0.10 reported by \citet{chun}.
However, if any of the brighter, blue candidates are confirmed as RGB members,
the TRGB magnitude would move brightward by $>$0.1 mag in all three filters.       

\noindent\textbf{NGC 6637 (M69):} Our CMD-selected TRGB candidate is nearly 4
$\arcmin$ ($\sim$4.7$R_{hl}$) 
from the cluster centre and its membership status could therefore be
considered uncertain.  If in fact a nonmember, 
the next brightest candidate after eliminating the 
large amplitude variables V1 and V3 is V7, which, if an RGB (not AGB)
variable, would shift the TRGB magnitudes faintward by $>$0.2 mag.

\noindent\textbf{NGC 6638:} We adopt the same tip star as \citet{v10}.  Although this is variable V70 of \citet{skottfelt}, it is likely an RGB rather than AGB variable given the period and amplitude they report.

\noindent\textbf{NGC 6642:} This is another case of some ambiguity in determination of
the TRGB magnitudes.  We have selected the same TRGB star as \citet{v10}, which is a likely member
at $<$$R_{hl}$.  However, after eliminating the Mira variable V2578 Sgr, there
is one significantly brighter ($>$0.4 mag in all three filters) candidate
which survived our decontamination procedure, although it lies much further
from the cluster center ($\sim$2.4$R_{hl}$).  Similarly, if our selected TRGB
candidate turns out to be a non-member or AGB star, there is another candidate at $R<R_{c}$ which is only 0.05 mag fainter in $K_{S}$ but $>$0.3 mag fainter in $J$ and $H$.

\noindent\textbf{NGC 6656 (M22):} In this case, there are two likely TRGB candidates with very similar photometry, both of which are confirmed members \citep{cudworthm22}.  The brighter of the two is V9, which has been found to be periodic (with a period and amplitude compatible with RGB status) but more recently appeared to be in a quiescent phase \citet{clement,agbm22}.  The TRGB candidate we adopt is 
the same tip star selected by \citet{monaco}, which is $<$0.05 mag fainter than V9 in $JHK_{S}$ and is not a known variable.  However, since this cluster is nearby
and has a large core, we cannot exclude the possibility that shallow
high spatial resolution imaging could reveal additional TRGB candidates,
and we note that the 2MASS-PSC gives Xflg=2 for this star.      

\vspace{\baselineskip}
\noindent Clusters from \citet{cohenispi}:
\vspace{\baselineskip}

\noindent\textbf{NGC 104 (47 Tuc):} The brightest candidate TRGB star is
variable LW5 from \citet{agb47tuc}, who claim a period of 74d superimposed on a more long-term variation.  Given this period and the relatively small amplitude shown in their fig.~1, this could be an RGB star, although an AGB status cannot be excluded based on its variability.

\noindent\textbf{NGC 288:} The tip star chosen both here and by \citet{v04abs} is a known
semi-regular variable with a $V$ amplitude of $A_{V}$=0.22 mag and a period of 103 days \citep{af288}.  
However, the next
brightest non-variable star with a $(J-K_{S})$ colour consistent with the
location of the RGB
is more than 0.5 (0.7) mag fainter in $J$ ($K_{S}$).

\noindent\textbf{NGC 362:} Selection of the TRGB location in this cluster is complicated by the
presence of several low amplitude variables near the RGB tip.  The brightest
candidate 
is variable LW6 from \citet{agb362} (=Sz56), which
appears to lie slightly redward of the cluster RGB but is only 16$\arcsec$
($\sim$0.3$R_{hl}$) 
from the cluster center.  Given its short period (34d) and small amplitude ($A_{V}$=0.075)
we consider this to be a likely RGB, rather than AGB star, noting that the brightest
non-variable star lies $>$0.4 mag faintward in $J$ and $K_{S}$.

\noindent\textbf{NGC 1261:} The selected tip star lies 18$\arcsec$ ($<$0.5$R_{hl}$) 
from the cluster centre and is therefore a likely member.  $H$-band photometry has been taken from 2MASS.

\noindent\textbf{NGC 1851:} The brightest candidate tip star in our catalog, with
($J,K_{S}$)=(9.797,8.704) is 5$\arcsec$ from the cluster centre and therefore
likely subject to blending.  The next best candidate based 
on $(J-K_{S}$) colour is V9, for which \citet{layden1851} report $A_{V}$=0.43 and
a period of $\sim$141 days.  Although this star has a sufficiently small amplitude that we 
cannot exclude an RGB status, in light of its relatively long period we instead adopt the
next faintest candidate, which is $<$0.04 mag fainter in $J$ and $K_{S}$ and
not known to be variable (again employing $J,K_{S}$ photometry from our ISPI
catalog and $H$ from 2MASS).  This star is also a likely cluster member since
it is 23$\arcsec$ ($\sim$0.75$R_{hl}$) from the cluster center.

\noindent\textbf{NGC 2808:} We exclude the brightest candidate TRGB star, variable V45, based on its long period (332d) and large amplitude ($A_{V}$=0.8).  The next faintest candidate is also a variable, V31, but we consider this a viable candidate RGB star given the shorter period (60d) and smaller amplitude ($A_{V}$=0.5).  It is almost certainly a member, given its distance from the cluster center of $\sim$0.9$R_{hl}$ as well as stellar parameters and abundances from high-resolution spectroscopy \citep{carretta2808}.

\noindent\textbf{NGC 4833:} We have rejected the two brightest candidate TRGB stars
based on their blue colours, although one of these is variable V9 with a period of 87.7d and
unknown amplitude, and therefore an RGB status cannot be completely ruled out. 
The next best candidate, 2MASS J12594940-7054147, has photometry in excellent agreement with the location of the upper RGB and is located at $<$0.9$R_{hl}$ from the cluster center. However, if either of the aforementioned brighter, bluer candidates is an RGB member, the TRGB 
magnitudes would move brightward by $<$0.16 and 0.08 mag in $J$ and $K_{S}$
respectively. 
       
\noindent\textbf{NGC 5927:} The chosen tip star is 1.08$\arcmin$ ($\sim$$R_{hl}$) from the 
cluster center, so a likely member, although given its slightly blue colour and the field contamination in the direction of this cluster, its membership remains to be confirmed.

\noindent\textbf{NGC 6304:} The brightest candidate, 2MASS J17145274-2927586, has not been
chosen as its relatively large distance from the cluster center
($>$3.1$R_{hl}$) implies that it may not be a member in light of the field
contamination towards this cluster.  We choose the next brightest candidate,
which is 0.25 and 0.06 mag fainter in $J$ and $K_{S}$, although at
1.35$R_{hl}$ from the cluster center, its membership also remains to be
confirmed.  This is variable V15, listed as NSV 08361 in \citet{samus},
although the \citet{clement} catalog states that it may be constant based on
the study of \citet{hartwick6304}.  The next two fainter candidates are well
within $R_{hl}$, and if our chosen candidate turns out to be a nonmember, a
confirmation of membership for these two latter candidates would move the TRGB
as much as $\sim$0.12 mag faintward in $J$ and $K_{S}$.      

\noindent\textbf{NGC 6496:} We exclude the brightest candidate, which is variable V7 in
the \citet{clement} catalog, as it lies near the cluster tidal radius and has
mean optical colours and magnitudes inconsistent with RGB membership
\citep{abbas6496}, in addition to a relatively blue colour from 2MASS.  Our
chosen TRGB star is the brightest of the next three best candidates, which are
all variables with mean optical photometry from \citet{abbas6496} as well as
single-epoch photometry from 2MASS placing them on the cluster RGB.  However,
we note that our selected TRGB candidate V4 has Xflg=2 in 2MASS.

\noindent\textbf{NGC 6584:} The chosen tip star is $>$0.6$R_{hl}$ from the cluster center ($\sim$0.6$R_{t}$) and therefore its membership status should be confirmed.  If a nonmember, the two next best candidates are both within the cluster $R_{hl}$ and would imply a faintward shift in the TRGB magnitude of 0.14 and 0.24 mag in $J$ and $K_{S}$ respectively.

\noindent\textbf{NGC 7099 (M 30):} We select the same TRGB star as
\citet{v04abs}, which is within the cluster $R_{hl}$.  While there are two
brighter candidates within the cluster tidal radius, they lie far
($>$12$R_{hl}$) from the cluster center, and in both cases their foreground
nature is confirmed by spectroscopic abundances \citep{rave} and distances \citep{ammons}. 
     
\vspace{\baselineskip}
\noindent Additional clusters from \citet{v04abs,v10}:
\vspace{\baselineskip}

\noindent\textbf{NGC 5904 (M 5):} 
We adopt the same TRGB star as \citet{v04abs}, variable V50.  \citet{af5904}
have determined a period of 107.6d, but given its small amplitude evident in 
their fig.~3 ($A_{I}$$\sim$0.3), 
its location in optical and near-IR CMDs, and its proximity to the cluster ($<$1.1$R_{hl}$) 
this star is a likely member.  However, if not a member, the next brightest candidate, which
lies with $R_{hl}$, would move the TRGB magnitude 0.14 and 0.1 mag faintward in $J$ and $K_{S}$.

\noindent\textbf{NGC 6205 (M 13):} The tip star we adopt, the same employed by \citet{v04abs}, is V24.  Given its period of 45.34 days and small amplitude ($A_{V}$=0.24) from the \citet{clement} catalog, its pulsational properties may be consistent with an RGB rather than AGB status.  
Alternatively, a significant ($\Delta$$Y$$>$0.05) 
helium enhancement in this cluster 
\citep[e.g.][]{caloi,johnsonm13,uvhbhe,v13} could
substantially affect the location of both the RGB bump and tip.

\noindent\textbf{NGC 6388:} The previously employed TRGB candidate \citep{v10}
is now known to be V3, a Mira variable \citep{sloanmira}.  Meanwhile, V12 
(=star 1 in the catalog of \citet{v10}) is a brighter candidate, although
it is listed as a long period variable with an amplitude of $A_{V}$=0.6,
leaving its evolutionary status uncertain.  However, in the Washington CMD of
\citet{hughes6388}, the location of this star appears inconsistent with the
location of the RGB, so we discard it.  As there are no
other viable TRGB candidates within $\sim$12$R_{hl}$, we therefore select the
next faintest candidate, which has Washington photometry placing it on
the RGB, and is not a known variable.  This represents a faintward shift of
$\sim$0.1 mag in $K_{S}$ and a \textit{brightward} shift of $>$0.2 mag in $H$
from the \citet{v10} candidate, and these shifts would be even larger (in
absolute value) if V12 were to be confirmed as an RGB, rather than AGB,
member.  
 
\noindent\textbf{NGC 6752:} We select the same tip star as \citet{v04abs},
2MASS J19110813-6001517, which is 3.43$\arcmin$ ($\sim$1.8$R_{hl}$) from the 
cluster center.  Although there are several brighter candidate tip stars, 
they lie significantly farther ($\gtrsim$5$R_{hl}$) from the cluster center,
and unfortunately none of these could be matched to recent proper motion or radial
velocity studies.

\bsp
\label{lastpage}
\end{document}